\documentclass[11pt,a4paper]{article}

\usepackage{eurosym}
\usepackage{booktabs}
\usepackage{multirow}
\usepackage{sectsty}
\usepackage{graphicx}
\usepackage{amsmath}
\usepackage{amssymb}
\usepackage{subcaption}
\usepackage{caption}
\usepackage{todonotes}
\usepackage{hyperref}
\hypersetup{colorlinks=true,linkcolor=blue,urlcolor=blue}
\usepackage{algorithm}
\usepackage{algpseudocode}
\algnewcommand\MyAnd{\textbf{and} }
\usepackage{float}
\newfloat{algorithm}{t}{lop}
\usepackage{listings}
\definecolor{dkgreen}{rgb}{0,0.6,0}
\definecolor{gray}{rgb}{0.5,0.5,0.5}
\definecolor{mauve}{rgb}{0.58,0,0.82}

\usepackage{fancyhdr}
\usepackage[left=2cm,top=1.5cm,bottom=1.5cm,right=2cm,includefoot,includehead,headheight=13.6pt]{geometry}
\usepackage{lastpage}
\usepackage{enumitem}
\usepackage{lipsum}
\usepackage{titlesec}
\usepackage[toc]{multitoc}

\usepackage{bm}
\numberwithin{equation}{section}

\newcommand{\hr}[4]{\bm{\langle\,}#1\bm{,}\, #2\bm{,}\, #3\bm{,}\, #4\bm{\,\rangle}}
\newcommand{\hrl}[3]{\bm{\langle\,}#1\bm{,}#2\bm{,}#3\bm{,}}
\newcommand{\hrr}[1]{#1\bm{\,\rangle}}

\newcommand{\ssfill}{\xleaders\hbox to 0.35em{\scriptsize.}\hfill}

\newcommand*{\cventry}[7][.25em]{
  \noindent\begin{tabular*}{\textwidth}{l@{\extracolsep{\fill}}r}%
	  {\bfseries #4} & {\bfseries #5} \\%
	  {\itshape #3\ifthenelse{\equal{#6}{}}{}{, #6}} & {\itshape #2}\\%
  \end{tabular*}%
  \ifx&#7&%
    \else{\\\vbox{\small#7}}\fi%
  \par\addvspace{#1}}

\newcommand*{\hintfont}{\bfseries}
\newcommand*{\hintstyle}[1]{{\noindent\hintfont{#1}}}
\newcommand*{\cvitem}[3][.25em]{%
  \ifthenelse{\equal{#2}{}}{}{\hintstyle{#2}: }{#3}%
  \par\addvspace{#1}}

\let\OriginalQuotation\quotation
\renewcommand*{\quotation}{\OriginalQuotation\small\sf}

\makeatletter
\g@addto@macro\bfseries{\boldmath}
\makeatother
\begin{document}

\allsectionsfont{\sffamily}

\fancyhead{}
\fancyfoot{}

\fancyhead[CO]{{}}
\fancyhead[LO]{{}}
\fancyhead[RO]{{}}
\fancyfoot[R]{\thepage\, / {\color[rgb]{0.6,0.,0}\pageref{LastPage}}}
\renewcommand{\headrulewidth}{0pt}

\newcommand*{\titleGP}{\begingroup 
\thispagestyle{empty}

{\centering 

\begin{center}
{\LARGE {\sf Automatic differentiation for error analysis of Monte
    Carlo data}} \\[0.2\baselineskip] 
\end{center}

}
\begin{center}
  {\large Alberto~Ramos} 
\end{center}
\vspace{0.2cm}
\begin{center}
  {School of Mathematics \& Hamilton Mathematics Institute, Trinity College Dublin, Dublin
  2, Ireland}\\
  \texttt{<alberto.ramos@maths.tcd.ie>}
\end{center}

\vspace{1cm}
\begin{center}
  \large{\sf Abstract}
\end{center}
\rule{\textwidth}{0.4pt}
\noindent
\emph{Automatic Differentiation} (AD) allows to determine exactly the
Taylor series of any function truncated at any order. Here we propose to
use AD techniques for Monte Carlo data analysis. We discuss how to
estimate errors of a general function of measured observables in
different Monte Carlo simulations. Our proposal
combines the $\Gamma$-method with Automatic differentiation, allowing
\emph{exact} error propagation in arbitrary observables, even those
defined via iterative algorithms. The case of special interest where
we estimate the error in fit parameters is discussed in detail. We
also present a freely available \texttt{fortran} reference
implementation of the ideas discussed in this work. 


\noindent\rule{\textwidth}{0.4pt}\\[\baselineskip] 

\tableofcontents

\newpage
\endgroup}

\begingroup 
\thispagestyle{empty}

{\centering 

\begin{center}
{\LARGE {\sf Automatic differentiation for error analysis of Monte
    Carlo data}} \\[0.2\baselineskip] 
\end{center}

}
\begin{center}
  {\large Alberto~Ramos} 
\end{center}
\vspace{0.2cm}
\begin{center}
  {School of Mathematics \& Hamilton Mathematics Institute, Trinity College Dublin, Dublin
  2, Ireland}\\
  \texttt{<alberto.ramos@maths.tcd.ie>}
\end{center}

\vspace{1cm}
\begin{center}
  \large{\sf Abstract}
\end{center}
\rule{\textwidth}{0.4pt}
\noindent
\emph{Automatic Differentiation} (AD) allows to determine exactly the
Taylor series of any function truncated at any order. Here we propose to
use AD techniques for Monte Carlo data analysis. We discuss how to
estimate errors of a general function of measured observables in
different Monte Carlo simulations. Our proposal
combines the $\Gamma$-method with Automatic differentiation, allowing
\emph{exact} error propagation in arbitrary observables, even those
defined via iterative algorithms. The case of special interest where
we estimate the error in fit parameters is discussed in detail. We
also present a freely available \texttt{fortran} reference
implementation of the ideas discussed in this work. 


\noindent\rule{\textwidth}{0.4pt}\\[\baselineskip] 

\tableofcontents

\newpage
\endgroup

\section{Introduction}

Monte Carlo (MC) simulations are becoming a fundamental source of
information for many research areas. In particular, obtaining first
principle predictions from QCD at low energies
requires to use Lattice QCD, which is based on the Monte Carlo sampling
of the QCD action in Euclidean space.
The main challenge when analyzing MC data is to
assess the statistical and systematic errors of a general complicated
function of the primary measured observables.

The autocorrelated nature
of the MC measurements (i.e. subsequent MC
measurements are not independent) makes error estimation difficult. The
popular resampling methods (bootstrap and jackknife) deal with the
autocorrelations of the data by first binning: blocks of data are
averaged in bins. It is clear that bins of data are less correlated
than the data itself, but the remaining correlations decrease slowly,
only inversely proportional with the size of the
bins~\cite{Wolff:2003sm}. The
$\Gamma$-method~\cite{Madras1988,Wolff:2003sm, Schaefer:2010hu} 
represents a step forward in the treatment of autocorrelations. Here
the autocorrelation function of the data is determined
explicitly and the truncation errors decay 
\emph{exponentially} fast at asymptotic large MC times, instead of
power-like~\cite{Wolff:2003sm}.  

In practice the situation is even more delicate. Large exponential
autocorrelation times are common in many current state of the art
lattice simulations and errors naively determined with either 
binning methods or the $\Gamma$-method might not even be in the asymptotic 
scaling region. The main difference between both approaches is that
the $\Gamma$-method allows to explicitly include an estimate of the
slow decay modes of the MC chain in the error
estimates~\cite{Schaefer:2010hu}. No similar techniques are available
for the case of binning methods.

A second issue in data analysis is how to properly assess an error to a
function of 
several MC observables, possibly coming from different ensembles. It
is important to note that this ``function'' can be a very non-linear
iterative procedure, like a fit or the solution of an iterative
method. Linear error propagation is the tool to determine the error in
these quantities, and requires the evaluation of the derivative of
this complicated function. Resampling methods compute these
derivatives stochastically by evaluating the non-linear function
for each sample and determining the standard deviation between the
samples of the function values. In the
$\Gamma$-method one usually evaluates the derivative by some finite
difference approximation. Both these approaches have some
drawbacks. First, they might not be computationally the most efficient
methods to propagate errors. More important is that
all finite difference formulae are ill-conditioned. In
resampling methods one might experience that the fit ``does not
converge'' for some samples. In the case of the $\Gamma$-method some
experience is required to choose the step size used for the numerical
differentiation.  

Alternatives to numerical differentiation are known. In particular
\emph{Automatic Differentiation}~\cite{wiki:ad} (AD) provides a very
convenient tool to perform the linear propagation of errors 
needed to apply the $\Gamma$-method to \emph{derived} observables. In
AD one determines the derivative of \emph{any} given function exactly
(up to machine precision). AD is based on the simple idea that any
function (even an iterative algorithm like a fitting procedure), is
just made of the basic operations and the evaluation of a few
intrinsic functions ($\sin, \cos, \dots$). In AD the differentiation
of each of these elementary operations and intrinsic functions is
hard-coded. Differentiation of complicated functions follows from the
decomposition in elementary differentials by the chain rule. As we
will see, AD is just the perfect tool for error propagation, and for a
robust and efficient implementation of the $\Gamma$-method for error
analysis.  

Much of the material presented here is hardly new. The $\Gamma$-method
has been the error analysis tool in the ALPHA collaboration for quite
some time and the details of the analysis of MC data have been
published in works~\cite{Wolff:2003sm, Schaefer:2010hu,Virotta2012Critical},
lectures~\cite{rainer-lottini:2014} and internal
notes~\cite{notes:error_prop}. The use of AD in linear  
propagation of errors is also not new. For example the \texttt{python}
package \texttt{uncertainties}~\cite{py-uncertainties} implements
linear propagation of errors using AD.

But the existing literature does not consider the general (and fairly
common) situation
where Monte Carlo data from simulations with different parameters enter the
determination of some quantity. Here we will consider this case, and
also analyze in detail the case of error propagation in iterative
algorithms. AD techniques allow to perform this task
\emph{exactly} and, as we will see, in some cases  
error propagation can be drastically simplified.  The
ubiquitous case of propagating 
errors to some fit parameters is one example: the
second derivative of the $\chi^2$ function at the minima is all that
is needed for error propagation. 
The numerical determination of this Hessian matrix by finite
difference methods is delicate since least squares problems are very
often close to ill-conditioned. We will see that with AD techniques
the Hessian is determined exactly, up to machine precision, providing
an \emph{exact} and fast alternative (the minimization of the $\chi^2$ is
only done once!) for error propagation in fit parameters.  

The implementation of the $\Gamma$-method is in general more involved 
than the error propagation with resampling. In order to compute the
autocorrelation function in an efficient way, the necessary
convolution has to be done using the Fast Fourier Transform. The
correct accounting of the correlation among different observables
requires bookkeeping the fluctuations for each Monte Carlo
chain. The available implementations~\cite{Wolff:2003sm,
  Schaefer:2010hu, DePalma:2017lww} do not consider the general
case of derived observables from simulations with different parameters\footnote{The
  standard analysis tool of the ALPHA collaboration, the 
  MATLAB package \texttt{dobs}~\cite{Virotta2012Critical,rainer-lottini:2014} deals with
  this general case, although the code is not publicly
  available.}. This might 
partially explain why this method is not very popular despite the
superior treatment of autocorrelations. Here we present
a freely available reference implementation~\cite{aderrors-mod}, that
we hope will serve to fill this gap. 

The paper is organized as follows. In section~\ref{sec:gamma} we
present a small 
review on analysis techniques, with emphasis on the $\Gamma$-method
and the advantages over methods based on
binning and/or resampling. Section~\ref{sec:ad} covers the topic of automatic
differentiation. We will explain how AD works, and focus on a
particular implementation suitable for error analysis. In
section~\ref{sec:fits} we explore the application of AD to the
analysis of MC data using the $\Gamma$-method, with special emphasis on
estimating errors in observables defined via iterative algorithms,
like fit parameters. 
In section~\ref{sec:worked-out-example} we show the analysis of some
Monte Carlo data with our techniques, compare it with the more
classical tools of error analysis, and comment the strong points of
the advocated approach. Appendix~\ref{sec:model} contains some useful
formulas on the exact error in a model where the 
autocorrelation function is a combination of decaying
exponentials. We also give explicit formulas for the errors using
binning techniques. Finally in appendix~\ref{sec:code} we introduce a
freely available implementation of the ideas discussed in this work.


\section{Analysis of Monte Carlo data}
\label{sec:gamma}

In this section we provide a small summary on different analysis
techniques for MC data, with special emphasis in the $\Gamma$-method.
We follow closely the
references~\cite{Wolff:2003sm, Virotta2012Critical, Schaefer:2010hu,
  notes:error_prop} with emphasis on the analysis of derived observables from
different ensembles~\cite{Virotta2012Critical,notes:error_prop}. The
reader should note that the main purpose of 
this work is not to compare different analysis techniques. We will
nevertheless comment on the advantages of the $\Gamma$-method over the
popular methods based on binning and resampling.  

\subsection{Description of the problem}

We are interested in the general situation where some primary
observables $A_i^\alpha$ are measured in several Monte Carlo
simulations. Here the index $\alpha$ labels the ensemble where the
observable is measured, while the index $i=1,\dots,N_{\rm obs}^\alpha$
runs over the observables measured in the ensemble $\alpha$.
Different ensembles are assumed to correspond to different
physical simulation parameters (i.e. different values of the lattice spacing or 
pion mass in the case of lattice QCD, different values of the
temperature for simulations of the Ising model, etc\dots).  

There are two generic situations that are omitted from the discussion
for the sake of the presentation (the notation easily becomes
cumbersome). First there is the case where different ensembles
simulate the same physical parameters but different algorithmic
parameters. This case is easily taken into account with basically the
same techniques as described here.  Second, there is the case where
simulations only differ in the seed of the random number generator
(i.e. different \emph{replica}). Replicas can be used to improve the
statistical precision and the error determination along the lines
described in~\cite{Wolff:2003sm}. We just point out that the available
implementation~\cite{aderrors-mod} supports both cases.

A concrete example in the context of simulations of the Ising model
would correspond to the following observables\footnote{Note that the
  numbering of observables does not need to be consistent across
  ensembles. How observables are labeled in each ensemble
  is entirely a matter of choice.}
\begin{eqnarray}
  A_1^1 = \langle \epsilon\rangle_{T_1};&\qquad&
  A_2^1 = \langle m\rangle_{T_1}\\
  A_1^2 = \langle m\rangle_{T_2}\,,
\end{eqnarray}
where $\epsilon$ is the energy per spin and $m$ the magnetization per
spin. The notation $\langle\cdot \rangle_{T}$ means that the
expectation value is taken at temperature $T$. 

In data analysis we are interested in determining the uncertainty in
derived observables $F \equiv f(A_i^\alpha)$ (i.e. functions of the
primary observables). For the case presented above a simple example of
a derived observable is
\begin{equation}
  \label{eq:exder}
  F = \frac{A_2^1 - A_1^2 }{A_1^1}  = \frac{\langle m\rangle_{T_1} - \langle m\rangle_{T_2} }{\langle
    \epsilon\rangle_{T_1}}\,. 
\end{equation}

A more realistic example of a derived observable in lattice QCD is the
value of the proton mass. This is a function of many measured primary observables
(pion, kaon and proton masses and possibly decay constants measured in
lattice units at several values of the lattice spacing and
volume). The final result (the physical proton mass) is a function of
these measured primary observables. Unlike the case of the example in
eq.~(\ref{eq:exder}), this function cannot be written explicitly: it
involves several fits to extrapolate the lattice data to the
continuum, infinite volume and the physical point (physical values of
the quark masses).  

Any analysis technique for lattice QCD must properly deal with the
correlations between the observables measured on the same ensembles
(i.e. $A_1^1=\langle \epsilon\rangle_{T_1}$ and $A_2^1=\langle
m\rangle_{T_1}$ in our first example), and with the autocorrelations
of the samples produced by any MC simulation. At the same time the
method has to be generic enough so that the error in complicated derived
observables determined by a fit or by any other iterative procedure
(like the example of the proton mass quoted above) can be properly
estimated. 

\subsection{The $\Gamma$-method}

In numerical applications we only have access to a finite set of MC
measurements for each primary observable $A^\alpha_i$ 
\begin{equation}
  \label{eq:1}
  a_i^\alpha(t)\,,\qquad t = 1, \dots, N_\alpha\,,
\end{equation}
where the argument $t$ labels the MC measurement, and the
number of measurements available in ensemble $\alpha$ is labeled by
$N_\alpha$ (notice that it is the same for all observables measured on
the ensemble $\alpha$). As estimates for the values $A_i^\alpha$ we
use the MC averages 
\begin{equation}
  \label{eq:2}
  \bar a_i^\alpha = \frac{1}{N_\alpha}\sum_{t=1}^{N_\alpha} a_i^\alpha(t)\,.
\end{equation}
For every observable we define the 
fluctuations over the mean in each ensemble  
\begin{equation}
  \delta_i^\alpha(t) = a_i^\alpha(t) - \bar a_i^\alpha\,.
\end{equation}

We are interested in determining the uncertainty in any derived
observable (i.e. a generic function of the primary observables)
\begin{equation}
  F \equiv f(A_i^\alpha)\,.
\end{equation}
The observable $F$ is estimated by
\begin{equation}
  \bar F = f(\bar a_i^\alpha)\,.
\end{equation}
In order to compute its error, we use linear propagation of errors
(i.e. a Taylor approximation of the function $f$ at $A_i^\alpha$)
\begin{equation}
  f(A_i^\alpha + \epsilon_i^\alpha) = F + f_i^\alpha \epsilon_i^\alpha + \mathcal O(\epsilon_i^2)\,. 
\end{equation}
where
\begin{equation}
f_i^\alpha = \partial_i^\alpha f|_{A_i^\alpha} = \frac{\partial
  f}{\partial A_i^\alpha}\Big|_{A_i^\alpha}\,.
\end{equation}
In practical situations these derivatives are evaluated at $\bar a_i^\alpha$
\begin{equation}
  \bar f_i^\alpha = \partial_i^\alpha f\Big|_{\bar a_i^\alpha}\,.
\end{equation}
We also need the
autocorrelation function of the primary observables. When estimated
from the own Monte Carlo data we use 
\begin{equation}
  \label{eq:gamma}
  \Gamma_{ij}^{\alpha\beta}(t) =
  \frac{\delta_{\alpha\beta}}{N_\alpha-t}\sum_{t'=1}^{N_\alpha-t} 
  \delta_i^\alpha(t+t')\delta_j^{\alpha}(t')\,.
\end{equation}

Finally, the error estimate for $F$ is given in terms of the
autocorrelation functions
\begin{equation}
  \label{eq:rho}
  \rho_F^{\alpha}(t) =
  \frac{\Gamma_F^{\alpha}(t)}{\Gamma_F^{\alpha}(0)}\,,\qquad
  \Gamma_F^\alpha (t) = \sum_{ij} \bar f_i^\alpha \bar f_j^\alpha \Gamma_{ij}^{\alpha \alpha}(t)\,,
\end{equation}
that are used to define the (per-ensemble) \emph{variances}
$\sigma_F^\alpha(F)$ and integrated autocorrelation times $\tau_{\rm
  int}^\alpha(F)$ given by
\begin{equation}
 \label{eq:tauint}
  (\sigma_F^\alpha)^2 = \Gamma_F^\alpha(0)\,, \qquad
  \tau_{\rm int}^\alpha(F) = \frac{1}{2} + \sum_{t=1}^\infty \frac{\Gamma_F^\alpha(t)}{\Gamma_F^\alpha(0)} \,.
\end{equation}
Since different ensembles are statistically uncorrelated, the total
error estimate comes from a combination in quadrature
\begin{equation}
  (\delta \bar F)^2 = \sum_\alpha \frac{(\sigma_F^\alpha)^2}{N_\alpha}
  2\tau_{\rm int}^\alpha(F)\,.
\end{equation}
In the $\Gamma$-method each ensemble is treated independently, which
allows to know how much each ensemble contributes to the error in
$F$. The quantity  
\begin{equation}
  \label{eq:error_contribution}
  R_\alpha(F) = \frac{(\sigma_F^\alpha)^22\tau_{\rm
      int}^\alpha(F)}{N_\alpha(\delta \bar F)^2} 
\end{equation}
represents the portion of the total squared error of $F$ coming from
ensemble $\alpha$.

A crucial step is to perform a truncation of the infinite sum in
eq.~(\ref{eq:tauint}). In practice we use 
\begin{equation}
  \tau_{\rm int}^\alpha(F) = \frac{1}{2} + \sum_{t=1}^{W^\alpha_F} \frac{\Gamma_F^\alpha(t)}{\Gamma_F^\alpha(0)}\,.
\end{equation}

Ideally $W^\alpha_F$ has to be large compared with the exponential
autocorrelation time ($\tau_{\rm exp}^\alpha$) of the simulation (the slowest
mode of the Markov operator). In 
this regime the truncation error is $\mathcal O(e^{-W^\alpha/\tau_{\rm
 exp}^\alpha})$. The problem is that the error in $\Gamma_F^\alpha(t)$
is approximately constant in $t$. At large
MC times the signal for $\Gamma_F^\alpha(t)$ is small, and a too large
summation window will just add noise to the determination of
$\tau_{\rm int}^\alpha$.

In~\cite{Wolff:2003sm} a practical recipe to choose
$W_F^\alpha$ is proposed based on minimizing the sum of the systematic
error of the truncation and the statistical error. In this proposal
it is assumed that $\tau_{\rm exp}^\alpha \approx S_\tau\tau_{\rm
  int}^\alpha$, where $S_\tau$ is a parameter that can be tuned by
inspecting the data\footnote{Values in the range $S_{\tau}\approx
  2-5$ are common.}.

\begin{figure}
  \centering
  \includegraphics[width=0.6\textwidth]{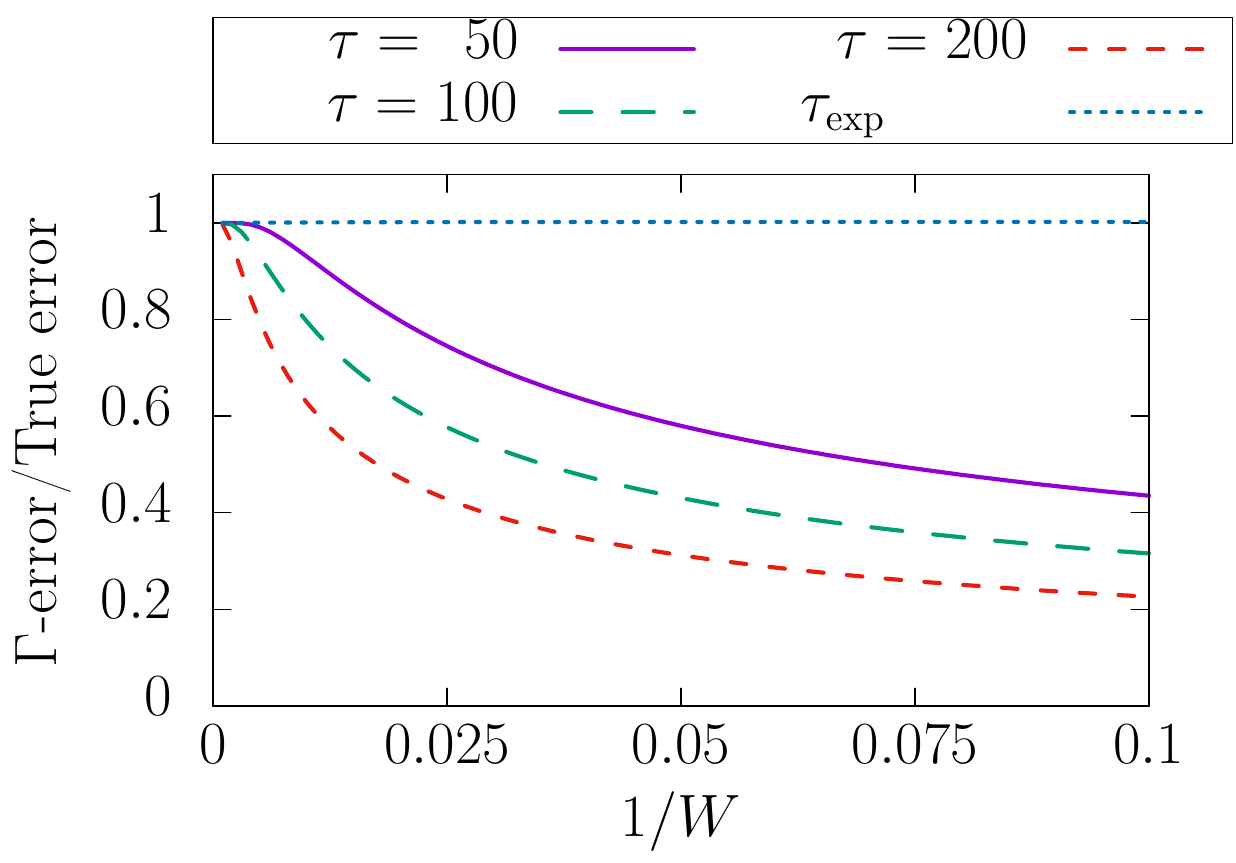}
  \caption{We show the approach to the true error in the
    $\Gamma$-method as a function of the summation window $W$ for
    three values of 
    $\tau$. In all cases the autocorrelation function is assumed to 
    follow a simple exponential decay $\Gamma(t) = e^{-|t|/\tau}$ (see
    appendix~\ref{sec:model}).  
    The asymptotic convergence is exponential, but in many
    practical situations we are far from this asymptotic
    behavior. As proposed in~\cite{Schaefer:2010hu} we can account for
    the slow modes in the analysis using eq.~(\ref{eq:tauint_tauexp})
    (corresponding to the line labeled $\tau_{\rm exp}$). In this
    very simple example (where the autocorrelation function has a single
    contribution) this procedure gives the correct error estimation
    for any window size and value of $\tau$.  
  }
  \label{fig:Wgamm} 
\end{figure}

Although this is a sensible choice, it has been
noted~\cite{Schaefer:2010hu} that in many practical situations in
lattice computations $\tau_{\rm exp}^\alpha$ and $\tau_{\rm
  int}^\alpha$ are in fact very different. In these cases the
summation window $W^\alpha_F$ cannot be taken much larger than $\tau^\alpha_{\rm
  exp}$, and one risks ending up underestimating the errors (see
figure~\ref{fig:Wgamm}). An improved 
estimate for $\tau_{\rm int}^\alpha$ was proposed for these
situations. First the autocorrelation function $\rho_F^\alpha(t)$ is
summed explicitly up to $W_F^\alpha$. This value has to be large, but
such that $\rho_F^\alpha(W_F^\alpha)$ is statistically different from
zero. For $t>W_F^\alpha$ we assume 
that the autocorrelation function is basically given by the
slowest mode $\rho_F^\alpha(t)\sim \exp(-|t|/\tau_{\rm exp}^\alpha)$
and explicitly add the remaining tail to the computation of $\tau_{\rm
  int}^\alpha$. The result of this operation can be summarized in the formula
\begin{equation}
  \label{eq:tauint_tauexp}
  \tau_{\rm int}^\alpha = \frac{1}{2} + \sum_{t=1}^{W^\alpha_F}
  \rho_F^\alpha(t) + \tau_{\rm exp}^\alpha\rho_F^\alpha(W_F^\alpha+1)\,.
\end{equation}

The original proposal~\cite{Schaefer:2010hu} consists in adding the
tail to the autocorrelation function at a point where $\rho_F(t)$ is
three standard deviations away from zero, and use the error
estimate~(\ref{eq:tauint_tauexp}) as an upper bound on the error. On
the other hand recent works of the ALPHA collaboration attach the tail
when the signal in $\rho_F(t)$ is about to be lost (i.e. $\rho_F(t)$
is 1-2 standard deviations away from zero) and use
eq.~(\ref{eq:tauint_tauexp}) as \emph{the} error estimate (\emph{not}
as an upper bound). This last option seems more appealing to the
author.  

All these procedures require an estimate of $\tau_{\rm
  exp}^\alpha$. This is usually obtained by inspecting large
statistics in cheaper simulations
(pure gauge, coarser lattice spacing, \dots). The interested reader is
invited to consult the original references~\cite{Schaefer:2010hu,
  Virotta2012Critical} for a full discussion.

\subsubsection{Notes on the practical implementation of the
  $\Gamma$-method}
\label{sec:notes-pract-impl}

In practical implementations of the $\Gamma$-method, as suggested
in~\cite{Wolff:2003sm},  it is convenient to 
store the mean and the \emph{projected fluctuations per
  ensemble} of an observable
\begin{equation}
  \label{eq:deltas}
  \bar F = f(\bar a_i^\alpha)\,,\qquad  \delta_F^\alpha(t) =
  \sum_{i}\bar f_i^\alpha \delta_i^\alpha(t)\,. 
\end{equation}
Note that observables that are functions of
other derived observables are easily analyzed. For
example, if we are interested in
\begin{equation}
  G = g(f(A_i^\alpha))\,,
\end{equation}
we first compute 
\begin{equation}
  \bar F = f(\bar a_i^\alpha)\,,\qquad \delta_F^\alpha(t) = \sum_{i}\bar f_i^\alpha \delta_i^\alpha(t)\,.
\end{equation}
Now to determine $G$ we only need an extra derivative
$\bar g_F = \partial_Fg\Big|_{\bar  F}$, since 
\begin{equation}
  \label{eq:error_propagation}
  \bar G = g(\bar F) = g(f(\bar a_i^\alpha))\,,\qquad  \delta_G^\alpha(t) =
\bar g_F\delta_F^\alpha(t) = \sum_{i}\bar g_F\bar f_i^\alpha
\delta_i^\alpha(t) = \sum_{i}\bar g_i^\alpha
\delta_i^\alpha(t)\,,
\end{equation}
with $\bar g_i^\alpha = \partial_i^\alpha g|_{\bar a_i^\alpha}$.

At this point the difference between a primary and a
derived observable is just convention: any primary observable can be
considered a derived observable defined with some identity
function. It is also clear that the means and the fluctuations are all
that is needed to implement linear propagation of errors in the
$\Gamma$-method.

Finally, we emphasize that computing derivatives of arbitrary
functions lies at the core of the
$\Gamma$-method. In~\cite{Wolff:2003sm, Schaefer:2010hu} a numerical
evaluation of the derivatives is used. Here we propose to use AD
techniques for reasons of efficiency and robustness, but before giving
details on AD we will comment on the differences between the
$\Gamma$-method and the popular methods based on binning and
resampling.

\subsection{Binning techniques}
\label{sec:comp-with-resampl}

\begin{figure}
  \centering
    \includegraphics[width=0.5\textwidth]{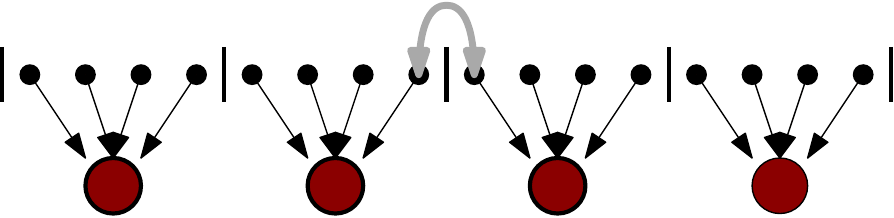}
    \caption{Blocks of data are averaged to produce a new data set with
    less autocorrelations. No matter how large the bins are,
    adjacent bins have always data that are very close in MC time.}
  \label{fig:bin}
\end{figure}

Resampling methods (bootstrap, jackknife) usually rely on binning to
reduce the autocorrelations of the data. One does not resample the
data itself, but bins of data. The original measurements
$a_i^\alpha(t)$ are first averaged in blocks of size $N_B$ (see
figure~\ref{fig:bin}) 
\begin{equation}
  b_i^\alpha(w) = \frac{1}{N_B}\sum_{t=1+(w-1)N_B}^{wN_B} a_i^\alpha(t)\,,
  \qquad (w=1,\dots,N_\alpha/N_B)\,.  
\end{equation}
This blocked data is then treated like independent measurements and
resampled, either with replacement in bootstrap techniques or by just
leaving out each observation (jackknife).

\begin{figure}
  \centering
  \includegraphics[width=0.6\textwidth]{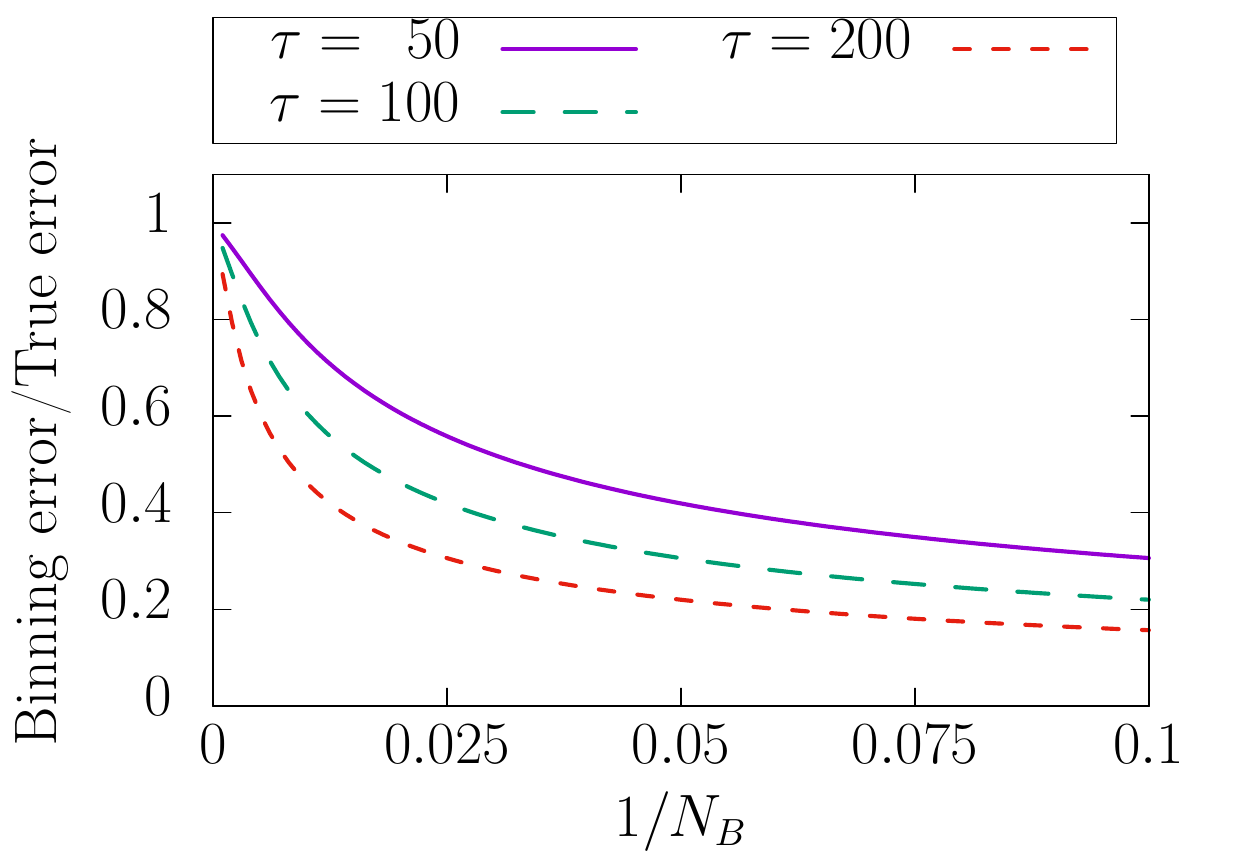}
  \caption{We show the approach to the true error by
    binning as a function of the bin size $N_B$ for three values of
    $\tau$. In all cases the autocorrelation function is assumed to 
    follow a simple exponential decay $\Gamma(t) = e^{-|t|/\tau}$ (see
    appendix~\ref{sec:model}). The asymptotic convergence is slow ($\mathcal
    O(1/N_B)$). Moreover in many practical situations we are far from this
    asymptotic behavior.}
  \label{fig:1ovN} 
\end{figure}

How good is the assumption that blocks are independent?
A way to measure this is to determine the autocorrelation function of
the blocked data (see
appendix~\ref{sec:ap_binning}). In~\cite{Wolff:2003sm} it is shown
that the leading 
term for large $N_B$ in the integrated autocorrelation time of the
binned data is $\mathcal O(1/N_B)$.
The autocorrelation function decays exponentially at large MC
time, but the fact that adjacent bins have always data that are very
close in MC time (see fig.~\ref{fig:1ovN}) transforms this expected
exponential suppression in a power law. 

In practice it is difficult to have bins of size much larger than
the exponential autocorrelation times. In this situation one is far
from the asymptotic $\mathcal O(1/N_B)$ scaling and binning severely
underestimates the errors. An instructive example is to just consider
a data with simple 
autocorrelation function $\Gamma(t) = e^{-|t|/\tau}$ (see
appendix~\ref{sec:ap_binning}). 
Figure~\ref{fig:1ovN} shows the approach to the true error as a
function of the bin size for different values of $\tau$. 

This example should be understood as a warning, and not as an academic
example: in state of the art lattice QCD simulations it is not uncommon
to simulate at parameter values where $\tau_{\rm exp}\sim 100-200$,
and it is fairly unusual to have statistics that allow for bins of
sizes larger than 50-100. We end up commenting that the situation in the
$\Gamma$-method is better for two reasons. First the truncation errors
are \emph{exponential} instead of power-like when enough data is
available. Second (and more important), in the cases where 
the statistics is not much larger than the exponential autocorrelation
times, an improved estimate like eq.~(\ref{eq:tauint_tauexp}) is only
available for the $\Gamma$-method.

\subsection{The $\Gamma$-bootstrap method}
\label{sec:resampl-gamma-meth}

Conceptually there is no need to use binning techniques with
resampling. Binning is only used to tame the autocorrelations in the
data, and as we have seen due to the current characteristics of the
lattice QCD simulations it seriously risks underestimating the errors.

Resampling is a tool for error propagation that automatically takes
into account the \emph{correlations} among different observables. A
possible analysis strategy consist in using the $\Gamma$-method to
determine the errors of the primary observables, and use resampling
techniques for error propagation.

The $N_{\rm p}\times N_{\rm p}$ covariance matrix among the primary
observables $A_i^\alpha$ can be estimated by
\begin{equation}
  \label{eq:cov}
  {\rm cov}(A_i^\alpha, A_j^\beta) \approx
  \frac{1}{N_\alpha}
  \left[
    \Gamma_{ij}^{\alpha\beta}(0) +
    2\sum_{t=1}^{W^\alpha}\Gamma_{ij}^{\alpha\beta}(t)
  \right]\,.
\end{equation}
The effect of large tails in the autocorrelation functions can also be
accounted for in the determination of this covariance matrix by using 
\begin{equation}
  \label{eq:cov_texp}
  {\rm cov}(A_i^\alpha, A_j^\beta) \approx
  \frac{1}{N_\alpha}
  \left[
    \Gamma_{ij}^{\alpha\beta}(0) +
    2\sum_{t=1}^{W^\alpha}\Gamma_{ij}^{\alpha\beta}(t) +
    2\tau_{\rm exp}^\alpha\Gamma_{ij}^{\alpha\beta}(W^\alpha+1)
  \right]\,.
\end{equation}
In the previous formulas the window $W^\alpha$ can be chosen with
similar criteria as described in section~\ref{sec:gamma}. Different
diagonal entries $i=j$ might give different values for
the window $W^\alpha_i$. For the case of eq.~(\ref{eq:cov}), the
values $W_i^{\alpha}$ are chosen with the criteria described
in~\cite{Wolff:2003sm}, and conservative error estimates are obtained
by using 
\begin{equation}
  W^\alpha = \max_i\{W_i^\alpha\}\,.
\end{equation}
On the other hand for the case of eq.~(\ref{eq:cov_texp}) we choose
$W_i^\alpha$ according to the criteria discussed before
eq.~(\ref{eq:tauint_tauexp}), and we use
\begin{equation}
  W^\alpha = \min_i\{W_i^\alpha\}\,.
\end{equation}

Once the covariance matrix is known, one can generate bootstrap
samples following a multivariate Gaussian distribution with the mean
of the observables $\bar a_i^\alpha$ as mean, and the covariance
among observables as covariance. These bootstrap samples are
used for the analysis as in any resampling analysis where the samples
come from binning. We will refer to this analysis technique as
$\Gamma$-bootstrap. 

It is clear that if each primary observable is measured in a different
ensemble the covariance matrix ${\rm cov}(A_i^\alpha,A_j^\beta)$ is diagonal
(cf. equation~(\ref{eq:gamma})). In this case the 
bootstrap samples are generated by just generating independent random
samples $\mathcal N(\bar a_i^\alpha, \sigma^2(a_i^\alpha))$ for each
observable. Moreover in this particular case the analysis of the data
will give completely equivalent results as using the
$\Gamma$-method. This is more clearly seen by looking at
equation~(\ref{eq:error_propagation}): any derived observable will
have just the same set of autocorrelation functions except for a
different scaling factor that enters the error
determination in a trivial way. 

\begin{table}
  \centering
  \begin{tabular}{lllllll}
    \toprule
    &&&& \multicolumn{3}{c}{Binning error/True error} \\
    \cmidrule(lr){5-7}
    Observable& Value & $\lambda_k$ & $\tau_{\rm int}$ & $N_B=10$ &$N_B=25$ &$N_B=50$ \\
    \cmidrule(lr){1-1}\cmidrule(lr){2-2}\cmidrule(lr){3-3}\cmidrule(lr){4-4}\cmidrule(lr){5-5}\cmidrule(lr){6-6}\cmidrule(lr){7-7}
    $x$ & 2.00(7) & $(1.08, 0.08, 0.05, 0.0)$ &4.5 &0.75 & 0.87 & 0.91 \\
    $y$ & 1.86(8) & $(1.00, 0.15, 0.0, 0.05)$ &6.1 &0.65 & 0.76 & 0.82 \\
    $z=x/y$ & 1.075(14) & $(0.0025,  -0.044, 0.027,  -0.029)$ &56.1 & 0.25 & 0.36 & 0.48\\
    \bottomrule
  \end{tabular}
  \caption{Large autocorrelation times can show up on derived
    observables even in cases where the primary observables have all
    small autocorrelations. The error in the column labeled ``value''
    shows the exact value of the observable and an error computed
    assuming a sample of length 2000. The values of $\lambda_k$,
    together with $\tau_k = (4.0, 100.0, 2.0, 3.0)$ give the
    autocorrelation functions $\Gamma(t) =
    \sum_k\lambda_k^2e^{-|t|/\tau_k}$ (see appendix~\ref{sec:model}). 
    }
  \label{tab:ratio}
\end{table}

But when several observables determined on the same ensembles
enter in the analysis, results based on this $\Gamma$-bootstrap
approach are not equivalent to analysis based on the
$\Gamma$-method. In resampling 
methods we loose all information on autocorrelations when we build the
bootstrap samples. When combining different observables from the same
ensembles the slow modes of the MC chain can be enhanced. Large
autocorrelations may show up in derived observables even in cases when
the primary observables have all small integrated autocorrelation
times. Table~\ref{tab:ratio} shows an example where the ratio of two
primary observables shows large autocorrelation times ($\tau_{\rm
  int}=56$) even in the case where the two primary observables have
relatively small autocorrelation times ($\tau_{\rm int}=4.5$ and
$\tau_{\rm   int}=6.1$). Note also that 
the derived observable is very precise (the error
is 6 times smaller than those of the primary observables).

Once more this example should not be considered as an academic
example\footnote{In general terms large autocorrelations tend to be
  seen in very precise data. Uncorrelated noise reduces autocorrelation
  times. A trivial example is to note that adding a white uncorrelated
  noise to any data decreases the integrated autocorrelation time of
  the observable (but of course \emph{not} the error!).}. Profiting
from the correlations among observables to obtain precise results lies
at the heart of many state of the art computations. As examples we can 
consider the ratio of decay constants ($F_K/F_\pi$ plays a central
role in constraining CKM matrix elements for example) or the
determination of isospin breaking effects. In these 
cases there is always the danger that the precise derived observable
shows larger autocorrelations than the primary observables. 

While one loses information on the autocorrelations for
derived observables (only a full analysis with the $\Gamma$-method has
access to this information), the large truncation errors expected
from binning are avoided by using a combination of the $\Gamma$-method
and resampling techniques for error propagation. Therefore this
analysis technique should always be preferred over analysis using
binning techniques.


\section{Automatic differentiation}
\label{sec:ad}

\label{sec:autom-diff}

At the heart of linear error propagation with the $\Gamma$-method lies
the computation of derivatives of arbitrary functions (see
eq.~(\ref{eq:error_propagation})). In this work we propose to use
Automatic differentiation (AD) to compute these derivatives. 

AD is a set of techniques to evaluate derivatives of functions.
A delicate point in numerical differentiation is the choice of step
size. If chosen too small, one will get large round-off errors. One
can also incur in systematic errors if the step size used is too
large. AD is free both of systematic and round-off errors:
derivatives of arbitrary functions are computed to machine
precision. 
At the core of the idea of using AD to compute derivatives lies the
fact that any function, even complicated iterative algorithms, are
just a series of fundamental operations and the evaluation of a few
intrinsic functions. In AD the derivatives of these fundamental
operations and few intrinsic functions are hard-coded, and the
derivatives of complicated functions follow from recursively applying
the basic rules.

In AD the decomposition of a derivative in derivatives of the
elementary operations/functions is fundamental. A central role is 
played by the chain rule. If we imagine a simple composition of functions
\begin{equation}
  y = f_3(f_2(f_1(x)))\,,
\end{equation}
and we are interested in the derivative ${\rm d}y/{\rm d}x$ we can define the following intermediate variables
\begin{equation}
  z_0 = x, \quad z_1 = f_1(x), \quad z_2 = f_2(z_1), \quad z_3 = f_3(z_2)\,.
\end{equation}
The chain rule gives for the derivative ${\rm d}y/{\rm d}x$ the
following expression
\begin{equation}
  \frac{{\rm d}y}{{\rm d}x} = \frac{{\rm d} y}{{\rm d} z_2} \frac{{\rm
      d} z_2}{{\rm d} z_1} \frac{{\rm d} z_1}{{\rm d} z_0}\,.
\end{equation}

AD can be applied to compute the previous derivative by using two
modes: \emph{reverse accumulation} and \emph{forward accumulation}.
In reverse accumulation~\cite{ad-reverse}, one starts the chain rule
with the variable that is being differentiated ($y$ in the previous
example), and proceeds to compute the derivatives
recursively.  
On the other hand, forward accumulation  determines the derivatives
with respect to the 
independent variable first ($x$ in our example). The  reader
interested in the details is invited to consult the literature on the
subject~\cite{griewank2008evaluating}.

Although more efficient for some particular problems,
the implementation of reverse accumulation AD is usually 
involved (see~\cite{griewank2008evaluating} for a full
discussion). On the other hand the implementation of forward
accumulation is more straightforward and it is naturally implemented
by overloading the fundamental operations and functions in languages
that support such features. 
For the particular applications described in this work,
most of the time of the analysis is spent in computing the
projected fluctuations eq.~(\ref{eq:deltas}) and in computing 
the autocorrelation function eq.~(\ref{eq:gamma}), therefore the
particular flavor of AD used does not influence the efficiency of the
analysis code. With these general points in mind, in the following
sections we will introduce a particularly convenient and simple
implementation of forward accumulation AD~\cite{Fike_HD}, that is
suitable for all the applications described in this work. 

\subsection{Forward accumulation AD and hyper-dual numbers}

In forward accumulation, the order for evaluating derivatives corresponds to the
natural order in which the expression is evaluated. This just asks to
be implemented by extending the operations ($+,-,*,/,\dots$) and
intrinsic functions ($\sin, \cos, \exp,\dots$) from the domain of the
real numbers. 

Hyper-dual numbers~\cite{Fike_HD} are represented as 4-components
vectors of real numbers $\tilde x = \hr{x_1}{x_2}{x_3}{x_4}$. They
form a field with operations
\begin{eqnarray}
  \tilde x \pm \tilde y &=& \hr{x_1 \pm y_1}{x_2 \pm y_2}{x_3 \pm y_3}
                            {x_4\pm y_4}\,, \\
  \tilde x \tilde y &=& \hr{x_1y_1}{x_1y_2+x_2y_1}{x_1y_3+x_3y_1}{x_1y_4 +
                        x_4y_1 + x_2y_3 + x_3y_2}\,, \\
  \tilde{x}^{-1}  &=& \hr{1/x_1}{-x_2/x_1^2}{-x_3/x_1^2}{2x_2x_3/x_1^3 -
                      x_4/x_1^2}\,.
\end{eqnarray}
Arbitrary real functions $f:\mathbb R\to \mathbb R$ are promoted to
hyper-dual functions by the relation 
\begin{equation}
  \tilde f(\tilde x) = \hr{f(x_1)}{f'(x_1)x_2}{f'(x_1)x_3}{x_4f'(x_1) + x_2x_3f''(x_1)}\,.
\end{equation}

Note that the function and its derivatives are only evaluated at the first component of
the hyper-dual argument $x_1$. In contrast with the techniques of symbolic
differentiation, AD only gives the values of the derivatives of a
function in one specific point. We also stress that the usual field 
for real numbers is recovered for hyper-real numbers of the form
$\tilde x = \hr{x_1}{0}{0}{0}$ (i.e. the real numbers are a sub-field
of the hyper-dual field).

Performing \emph{any} series of operations in
the hyper-dual field, the derivatives of any expression are
automatically determined at the same time as the value of the
expression itself. It is instructive to explicitly check the case of a
simple function composition
\begin{equation}
  y=f(g(x))\,.
\end{equation}
If we evaluate this at the hyper-real argument $\tilde x$, it is
straightforward to check that one gets
\begin{eqnarray}
  \tilde z = \tilde g(\tilde x) &=& \hr{g(x_1)}{g'(x_1)x_2}{g'(x_1)x_3}{x_4g'(x_1) + x_2x_3g''(x_1)}\,.\\
  \tilde y = \tilde f(\tilde z) &=& \hr{f(z_1)}{f'(z_1)z_2}{f'(z_1)z_3}{z_4f'(z_1) + z_2z_3f''(z_1)} \\
           &=& \hrl{f(g(x_1))}{f'(g(x_1))g'(x_1)x_2}{f'(g(x_1))g'(x_1)x_3}\\
           &&    \hrr{f'(g(x_1))(x_4g'(x_1) + x_2x_3g''(x_1)) + 
               g'^2(x_1)f''(g(x_1))x_2x_3}\,.
\end{eqnarray}
Now if we set $\tilde x = \hr{x_1}{1}{1}{0}$, we get $y'$ as the second/third
component of $\tilde y$, and $y''$ as the fourth component of $\tilde y$. 

\subsection{Functions of several variables}

The extension to functions of several variables is
straightforward. Each of the real arguments of the function is
promoted to an hyper-dual number. The hyper-dual field allows the
computation of the gradient and the Hessian of arbitrary
functions.

For a function of several variables $f(\mathbf{x})$, after
promoting all its arguments to the hyper-dual field we get the
hyper-dual function $\tilde f(\tilde{\mathbf{x}})$. Partial derivatives of the
original function $\partial_if(\mathbf{x})$ and the Hessian
$\partial_i\partial_jf(\mathbf{x})$ can be obtained using appropriate
hyper-dual arguments (see table~\ref{tab:multi} for an explicit
example with a function of two variables).

\begin{table}
  \centering
  \begin{tabular}{lll}
    \toprule
    $\tilde x$& $\tilde y$& $\tilde f(\tilde x,\tilde y)$\\
    \midrule
    $\hr{x_1}{1}{1}{0}$ & $\hr{y_1}{0}{0}{0}$ &
                                                $\hr{f}{\partial_x
                                                f}{\partial_xf}{\partial_x^2f}$
    \\ 
    $\hr{x_1}{1}{0}{0}$ & $\hr{y_1}{0}{1}{0}$ &
                                                $\hr{f}{\partial_x
                                                f}{\partial_yf}{\partial_x\partial_yf}$ \\
    $\hr{x_1}{0}{0}{0}$ & $\hr{y_1}{1}{1}{0}$ &
                                                $\hr{f}{\partial_y
                                                f}{\partial_yf}{\partial_y^2f}$
    \\
    \bottomrule
  \end{tabular}
  \caption{Example output for a function of two variables $f(x,y)$
    after promoting the arguments to hyper-real numbers. In the third
    column the function and all derivatives are evaluated at $(x_1,y_1)$.}
  \label{tab:multi}
\end{table}


\section{Applications of AD to the analysis of MC data}
\label{sec:appl-ad-analys}

The common approach to error propagation in a generic function
consists in examining how the function behaves when the input is
modified within errors. For example in resampling techniques the
target function is evaluated for each sample, and the spread in the
function values is used as an estimate of the error in the value of the
function. This approach is also used in cases where the function is 
a complicated iterative procedure. \emph{The main} example being error
propagation in fit parameters, that is usually performed by repeating
the fit procedure for slightly modified values of the data points. 

AD techniques propose an interesting alternative to this general
procedure: just performing every operation 
in the field of hyper-dual numbers will
give the derivatives necessary for error propagation \emph{exactly},
even if we deal with a complicated iterative algorithm
(section~\ref{sec:exact-error-prop} contains an explicit example). This is
relatively cheap numerically. For example in the case of
functions of one variable the numerical cost is
roughly two times the cost of evaluating the function if one is
interested in the value of the function and its first derivative, and
between three and four times 
if one is interested in the first and second derivatives. This has to
be compared with the three evaluations that are needed to obtain
the value of the function and \emph{an estimate} of the first
derivative using a symmetric finite difference, or the $\mathcal
O(1000)$ evaluations that are used in a typical resampling approach. 

In the rest of this section we will see that in many iterative
algorithms error propagation can be simplified. The most interesting
example concerns the case of error 
propagation in fit parameters: we will see that it is enough to
compute the second derivative of the $\chi^2$ function at the
minima. As a warm up example, we will consider the simpler (but
also interesting) case of error propagation in the root of a
non-linear function.   

\subsection{Error propagation in the determination of the root of a function}
\label{sec:newton}

A simple example is finding a root of a non-linear function of one
real variable. 
We are interested in the case when
the function depends on some data $d_a\, (a=1,\dots,N_{\rm
  data})$ that are themselves Monte Carlo observables\footnote{In the
terminology of section~\ref{sec:gamma} $d_a$ are some primary or derived
observables. }, $f(x;d_a)$. In this case
the error in the MC data $d_a$ propagates into an error of the
root of the function. We assume that for the central values of the data $\bar
d_a$ the root is located at $\bar x$
\begin{equation}
  f(\bar x; \bar d_a) = 0\,.
\end{equation}
For error propagation we are interested in how much changes the
position of the root when we change the data. This ``derivative'' can
be easily computed. When we shift the data around its central value
$\bar d_a\to \bar d_a+\delta d_a$, to leading order the function changes to
\begin{equation}
  f(x;\bar d_a+\delta d_a) = f(x;\bar d_a) +
  \partial_a f\Big|_{(x;\bar d_a)}\delta d_a\qquad (\partial_a
  \equiv \partial/\partial d_a)\,.
\end{equation}
This function will no longer vanish at $\bar x$, but at a slightly
shifted value $x = \bar x+\delta x$. Again to leading order
\begin{equation}
  f(\bar x+\delta x;\bar d_a+\delta d_a) = f(\bar x;\bar d_a) +
  \partial_x f\Big|_{(\bar x;\bar d_a)}\delta x +  
  \partial_a f\Big|_{(\bar x;\bar d_a)}\delta d_a = 0\,.
\end{equation}
This allows to
obtain the derivative of the position of the root with respect to the
data
\begin{equation}
  \frac{\delta x}{\delta d_a} = - \frac{\partial_a f}{\partial_x
    f}\Big|_{(\bar x;\bar d_a)}\,.
\end{equation}
That is the quantity needed for error propagation. Note that in
practical applications the
iterative procedure required to find the root (i.e. Newton's method,
bisection, \dots)
is only used once (to find the position of the root). Error
propagation only needs the derivatives of the target function at
$(\bar x,\bar d_a)$. 

\subsection{Error propagation in fit parameters}
\label{sec:fits}

In (non-linear) least squares one is usually interested in finding the
values of some parameters $p_i\, (i=1,\dots,N_{\rm parm})$ that minimize
the function
\begin{equation}
  \label{eq:chisq}
  \chi^2(p_i;d_a)\,,\qquad p_i\, (i=1,\dots,N_{\rm parm})\,, \quad
  d_a\, (a=1,\dots,N_{\rm data})\,.
\end{equation}
Here $d_a$ are the data that is fitted. In many cases the
explicit form of the $\chi^2$ is
\begin{equation}
  \chi^2 = \sum_{a=1}^{N_{\rm data}} \left( \frac{f(x_a;p_i) -
      y_a}{\sigma(y_a)} \right)^2\,,
\end{equation}
where $f(x_a;p_i)$ is a function that depends on the parameters $p_i$,
and $y_a$ are the data points (represented by $\{d_a\}$
in eq.~(\ref{eq:chisq})). The result of the fit is some parameters
$\bar p_i$ that make 
$\chi^2(\bar p_i;\bar d_a)$ minimum for some fixed values of the data $\bar d_a$. 

When propagating errors in a fit we are interested in how much 
the parameters change when we change the data. This ``derivative'' is defined
by the implicit condition that the $\chi^2$ has to stay always at its
minimum. If we shift the data $d_a \to \bar d_a + \delta
d_a$ we have to leading order
\begin{equation}
  \chi^2(p_i;\bar d_a + \delta d_a) = \chi^2(p_i;\bar d_a) +
  \partial_a\chi^2\Big|_{(p_i;\bar d_a)}\, \delta d_a\,, \qquad
  (\partial_a \equiv \partial/\partial d_a)\,.
\end{equation}
this function will no longer have its
minimum at $\bar p_i$ but will be shifted by an amount $\delta
p_i$. Minimizing with respect to $p_i$ and expanding at $p_i=\bar p_i$ to leading
order one obtains the condition
\begin{equation}
  \partial_j \partial_i \chi^2\Big|_{(\bar p_i;\bar d_a)}\delta p_j + \partial_i
  \partial_a\chi^2\Big|_{(\bar p_i;\bar d_a)}\, \delta d_a = 0\,, \qquad
  (\partial_i \equiv \partial/\partial p_i)\,.
\end{equation}
Defining the Hessian of the $\chi^2$ at the minimum
\begin{equation}
  H_{ij} = \partial_j \partial_i \chi^2\Big|_{(\bar p_i;\bar d_a)}\,,
\end{equation}
we can obtain the derivative of the fit parameters with respect to the
data~\footnote{As with the case of the normal equations, the reader is advised to
implement the inverse of the Hessian using the \texttt{SVD}
decomposition to detect a possibly ill-conditioned Hessian matrix}
\begin{equation}
  \frac{\delta p_i}{\delta d_a} = -\sum_{j=1}^{N_{\rm
      parm}}(H^{-1})_{ij}\partial_j \partial_a \chi^2\Big|_{(\bar p_i;\bar d_a)}\,.
\end{equation}
Once more the iterative procedure is only used one time
(to find the central values of the fit parameters), and error
propagation is performed by just evaluating derivatives of the
$\chi^2$ function at $(\bar p_i, \bar d_a)$.


\section{Worked out example}
\label{sec:worked-out-example}

As an example of the analysis techniques described in the text we are
going to study a simple non-linear fit. We want to describe  the
functional form $\tilde y(x)$ in the region $x\in [1,5]$. For this
purpose we have measured five values of $\tilde y$ at five values of
$x=1.0,2.0,3.0,4.0,5.0$ respectively. We are going to assume that the
data is well described by the model
\begin{equation}
  \label{eq:4}
  y = \log(n + x) + \sin(mx)\,,\qquad(\text{with }y = Z\tilde y)\,.
\end{equation}
The factor $Z$ is also part of the available measurements. 
Once  the parameters $n$ and $m$ are determined by fitting
our data $(x_i,\tilde y_i)$ we will have a parametrization of the
function $\tilde y(x)$. Note that equation~(\ref{eq:4}) describes a
highly non-linear function, both in the fit parameters and the
independent variable $x$.   

In the terminology of section~\ref{sec:gamma} we have six primary
observables (the common factor $Z$ and the five $\tilde y_i$) measured
each of them in a different ensemble
\begin{equation}
  A_1^1 = \tilde y_1, \, A_1^2 = \tilde y_2, \, A_1^3=\tilde y_3,
  \, A_1^4=\tilde y_4, \, A_1^5=\tilde y_5, \, A_1^6=Z\,.
\end{equation}
The exact values of these primary observables together with the
parameters used to generate the MC data are described in
table~\ref{tab:primary}. As examples of quantities of interest, we
will focus on obtaining the values and uncertainties of
\begin{enumerate}
\item The fit parameters $n,m$.
\item The value of the fitted function at $x=3/2$
  \begin{equation}
    I = \log\left(\frac{3}{2} + n\right) + \sin\left(\frac{3m}{2}\right)
  \end{equation}
\end{enumerate}
Note that these quantities are derived observables in the terminology
of section~\ref{sec:gamma} (i.e. functions of the primary
observables, defined via the fitting procedure). 

\begin{table}
  \centering
  \begin{tabular}{lllllll}
    \toprule
    &$\tilde y_1$&$\tilde y_2$&$\tilde y_3$&$\tilde y_4$&$\tilde y_5$&$Z$\\
    \midrule
    Value & 0.9021 & 1.5133 & 1.9319 & 2.1741 & 2.2508 & 1.2 \\
    $\tau$ & 2& 4& 6& 8& 10& 100\\
    $\lambda$ & 1 & 1&1&1&1&0.1\\
    $\tau_{\rm int}$ & 2.0415& 4.0208& 6.0139& 8.0104& 10.008&
                                                               100.00\\
    \bottomrule
  \end{tabular}
  \caption{Primary observables of our worked out example. The row
    labeled Value shows the exact value of the observable. Rows $\tau$
  and $\lambda$ show the values of the parameters used to generate the
MC samples along the lines discussed in appendix~\ref{sec:model} (the
MC data follows a simple autocorrelation function $\sim
e^{-|t|/\tau}$). Row $\tau_{\rm int}$ shows the exact value of the
integrated autocorrelation time of the observable.}
  \label{tab:primary}
\end{table}

In order to estimate our derived observables, we have at our disposal
2000 MC measurements for each of the six primary observables.
Table~\ref{tab:data} shows the estimates of these six primary
observables using different analysis techniques: on one hand the
$\Gamma$-method, where we use the improved error estimate
(eq~(\ref{eq:tauint_tauexp})) with $\tau_{\rm exp}=100$ for the
analysis of the observable $Z$, and the usual automatic window
procedure described in~\cite{Wolff:2003sm} with $S_\tau=2.0$ for the
five observables $\tilde y_i$. On the other hand we use the more
common binning/resamplnig techniques with different bin 
sizes. 
\begin{table}
  \centering
  \begin{tabular}{lllllll}
    \toprule
    &\multicolumn{2}{c}{$\Gamma$-method} & \multicolumn{3}{c}{Binning} \\ 
    \cmidrule(lr){2-3} \cmidrule(lr){4-6}
    & & & {$N_{\rm bin} = 10$}
                                          & {$N_{\rm bin} = 25$} 
                                          & {$N_{\rm bin} = 50$} \\
    \cmidrule(lr){4-4}
    \cmidrule(lr){5-5} \cmidrule(lr){6-6} 
    Obs. & $\tau_{\rm exp}$ &&&&&Error\\
\midrule
 $\tilde y_1$  & 0 & 0.881(50) &  0.881(43) &  0.881(49) & 0.881(54)&0.045  \\
 $\tilde y_2$  & 0 & 1.448(47) &  1.448(47) &  1.448(45) & 1.448(46)&0.063 \\ 
 $\tilde y_3$  & 0 & 1.981(91) &  1.981(62) &  1.981(78) & 1.981(85)&0.078 \\
 $\tilde y_4$  & 0 &  2.21(11) &  2.206(57) &  2.206(72) & 2.206(93)&0.090 \\
 $\tilde y_5$  & 0 & 2.309(96) &  2.309(58) &  2.309(76) & 2.309(75)&0.100
    \\
    \midrule
 $Z$ &100&$1.188(40)$ &{$1.1878(75)$} & $1.188(12)$
                                                    & $1.188(16)$&0.032 \\
    \bottomrule
  \end{tabular}
  \caption{Estimates of the values of the primary observables $\tilde
    y_i, Z$ using
    different analysis techniques. The last column labeled ``error''
    quotes the exact error of the observable assuming a sample of
    length 2000 (see
    appendix~\ref{sec:model}). The analysis of the primary observables
    is performed with different techniques: the $\Gamma$-method, where
    the slow mode in observable $Z$ is taken into account as describd
    in eq.~(\ref{eq:tauint_tauexp}), and the more traditional
    binning/resampling where we have choosen different bin sizes.} 
  \label{tab:data}
\end{table}

The strategy to determine the derived observables $n,m,I$ is the usual
one: we fit our data by minimizing the $\chi^2$ function\footnote{Note that
  the values of $y_a$ are correlated because of the common factor $Z$
  in equation~(\ref{eq:4}). We are going to perform an
  \emph{uncorrelated} fit, but the correlations among the data
  are taken into account in the error propagation.}
\begin{equation}
    \chi^2 = \sum_{a=1}^{5} \left( \frac{\log(n + x_a) + \sin(mx_a) -
      y_a}{\sigma(y_a)} \right)^2\,,\qquad (y_a = Z\tilde y_a)\,.
\end{equation}

We will use different approaches to determine the fit
parameters and $I$. 
\begin{enumerate}
\item First we will use our proposal of section~\ref{sec:fits}:
  We use any fitting routine, and once the minimum is found, the Hessian
  is determined with AD techniques and linear error propagation is
  performed. We will perform the analysis including 
  the tail in the autocorrelations function of the slow observable
  $Z$ (with $\tau_{\rm exp}=100$). The
  tail will be added when $\rho(t)$ is 1.5 times its
  error. For the other
  observables we will just assume that we have enough data so that
  truncation effects have negligible systematics.

\item As is clear from the discussion in
  section~\ref{sec:comp-with-resampl} binning methods tend to
  underestimate the true error\footnote{This is also apparent
    looking at table~\ref{tab:data}, where a bin size of $50$ is
    needed in order not to underestimate data with $\tau_{\rm int}=10$. The
    error of the slow observable $Z$ is severely underestimated for all
    reasonable bin sizes.}. In section~\ref{sec:resampl-gamma-meth} we
  detailed the $\Gamma$-bootstrap method, where errors of primary
  observables are computed with the $\Gamma$-method and resampling
  is used for error propagation. Since observable $Z$ is a primary
  observable we can add a
  tail to the autocorrelation function using $\tau_{\rm exp}=100$. The
  tail will be added when $\rho(t)$ is 1.5 times its error.

\item Finally we have used the more common binning and resampling
  approach. We use bins of size 10, 25, 50 so that we are left with
  200, 80 and 40 measurement respectively. These measurements are
  resampled (we use 2000 samples) with replacement (bootstrap) in
  order to perform the usual error analysis. 
\end{enumerate}
We note however that comparing binning with the
$\Gamma$-method is not the main purpose of this section. A detailed
comparison requires to either push the approach outlined in
appendix~\ref{sec:model} to compare error determinations in this
model, or to repeat the same analysis with several data sets that only
differ in the random number seed. These comparisons are available in the
literature~\cite{Wolff:2003sm, Schaefer:2010hu,
  Virotta2012Critical}. Here we will focus on how to perform the
analysis with the $\Gamma$-method when different ensembles enter our
error determination and in the use of AD techniques.

The results of this small experiment are summarized in
table~\ref{tab:res}. As the reader can see the $\Gamma$-method
together with linear error propagation using AD techniques is much
more efficient in terms of computer time. It also leads to
conservative error estimates. Even in this mild case, where our
primary measured observables have most of its contributions coming
from fast Monte Carlo modes ($\tau\sim 2-10$), binning methods
severely underestimate the errors unless one has access to very large
bin sizes. This is of course expected on theoretical
grounds~\cite{Wolff:2003sm} (see appendix~\ref{sec:ap_binning}).
\begin{table}
  \centering
  \begin{tabular}{lllllll}
    \toprule
    &&$\Gamma$-method (AD) &$\Gamma$-bootstrap &
                                                                 \multicolumn{3}{c}{Binning
                                                                 + bootstrap} \\ 
    \cmidrule(lr){3-3} \cmidrule(lr){4-4} \cmidrule(lr){5-7}
    Qauntity & Value &\multicolumn{2}{c}{$\tau_{\rm exp} = 100$} & {$N_{\rm bin} = 10$}
                                          & {$N_{\rm bin} = 25$} 
                                                              & {$N_{\rm bin} = 50$} \\
    \cmidrule(lr){1-1} \cmidrule(lr){2-2} \cmidrule(lr){3-4}
    \cmidrule(lr){5-5} \cmidrule(lr){6-6} \cmidrule(lr){7-7}
    $n$ & 1.0 & 0.98(19)  & 0.98(21)  & 0.97(12)  &0.97(15)  &0.97(17) \\ 
    $m$ & 0.4 & 0.365(45) & 0.364(56) & 0.372(32) &0.367(41) &0.365(44)\\ 
    $I$ & 1.4809...& 1.429(62) & 1.428(61) & 1.435(35) &1.429(39) &1.425(43)\\ 
\cmidrule(lr){3-3}\cmidrule(lr){4-7}
    Time & & \multicolumn{1}{c}{1} & \multicolumn{4}{c}{$\sim 250$}\\
    \bottomrule
  \end{tabular}
  \caption{Results for the derived observables $n,m$ and $I$ using
    different analysis 
    methods. The column ``Value'' gives the exact value of the
    parameters. We show the estimates coming from different analysis
    techniques. Note that binning severely underestimates the errors
    unless one has access to very large bins of data. As expected for
    this case the mixed
    $\Gamma$-bootstrap approach for the error propagation
    shows no difference from the results obtained via the
    $\Gamma$-method (see the discussion on the text). Error propagation using AD
    techniques, where the fit is only performed once, is much more
    efficient in terms of computing time. (Times were measured by
    repeating the analysis 100-1000 times in a standard laptop. The
    timing does not pretend to be accurate, just to give an idea of
    the order of magnitude).}
  \label{tab:res}
\end{table}
The $\Gamma$-bootstrap method
(cf. section~\ref{sec:resampl-gamma-meth}) is a safe alternative to 
the full analysis using the $\Gamma$-method in this case. Note
that we have only one primary observable from each ensemble, and
therefore there is no possibility of any cancelation that would
uncover some large autocorrelations (see section~\ref{sec:resampl-gamma-meth}). 
The $\Gamma$-method is still faster due to the fact that resampling methods
have to perform the fit many times (2000 in the example above), while
AD techniques allow to perform the error propagation with only a
single minimization. On the other hand the $\Gamma$-method requires to
keep track of all the fluctuations per ensemble (in the example above
the fit parameters are derived observables with 2000 fluctuations for
each of the 6 primary observables) and perform some FFTs to do the
necessary convolutions needed to determine the autocorrelation
function. Still the analysis with the $\Gamma$-method is around two
orders of magnitude faster than resampling techniques.

\begin{figure}
  \centering
  \includegraphics[width=0.95\textwidth,height=0.15\textheight]{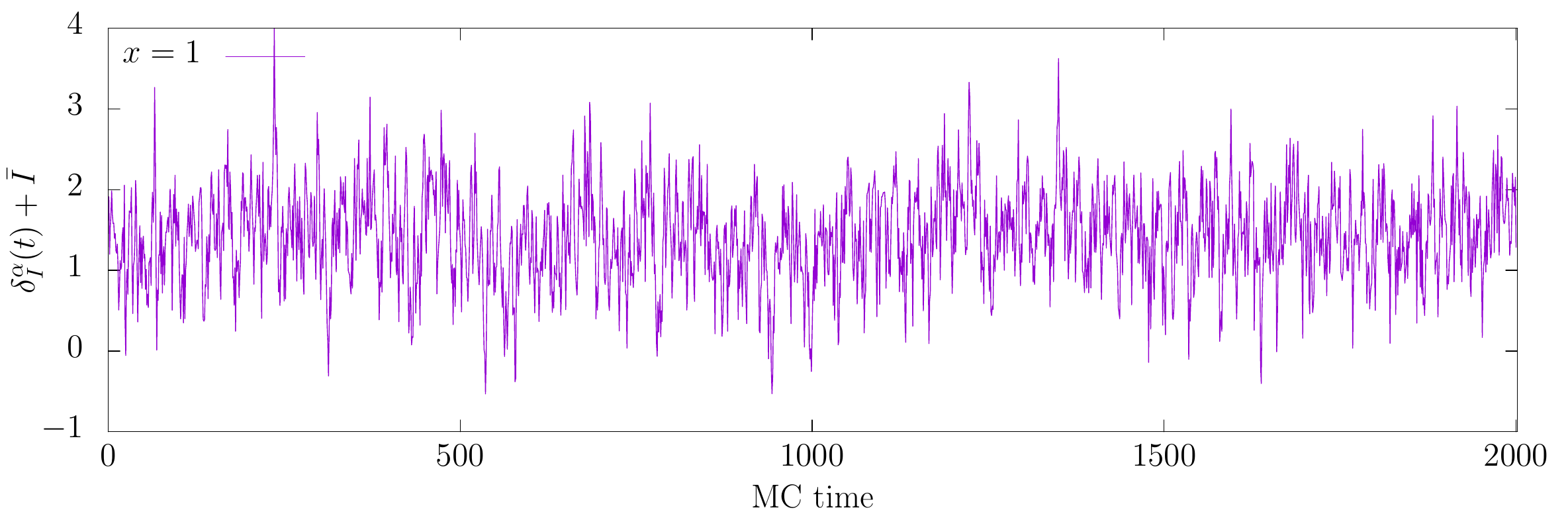}\\
  \includegraphics[width=0.95\textwidth,height=0.15\textheight]{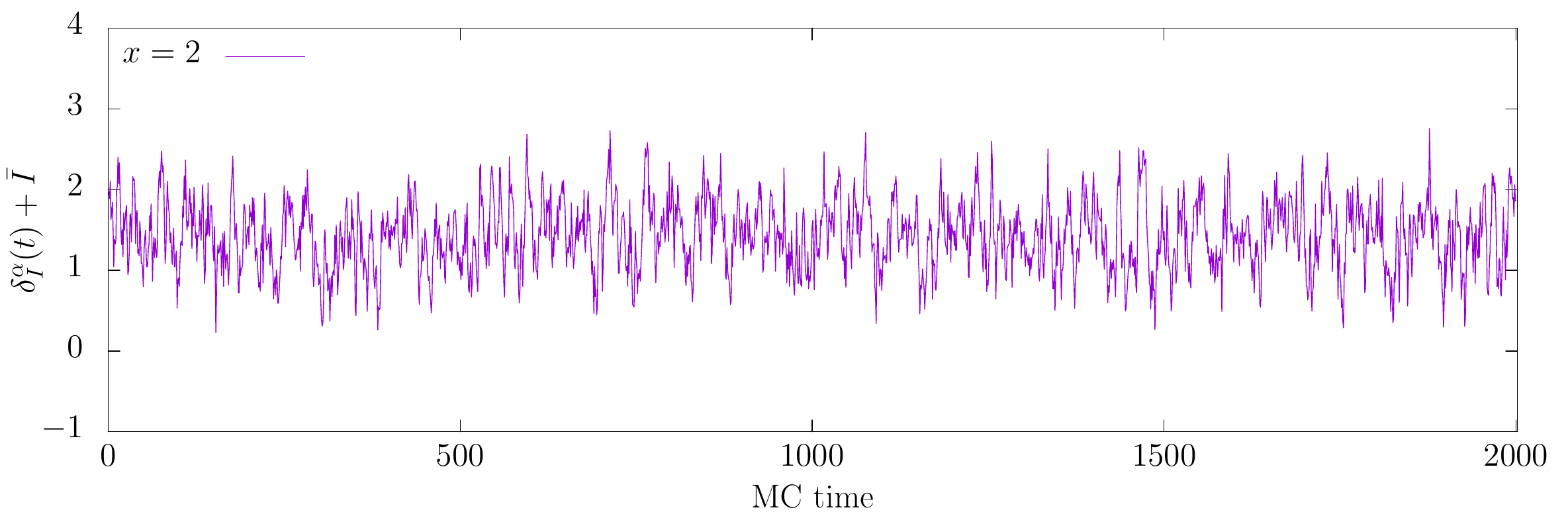}\\
  \includegraphics[width=0.95\textwidth,height=0.15\textheight]{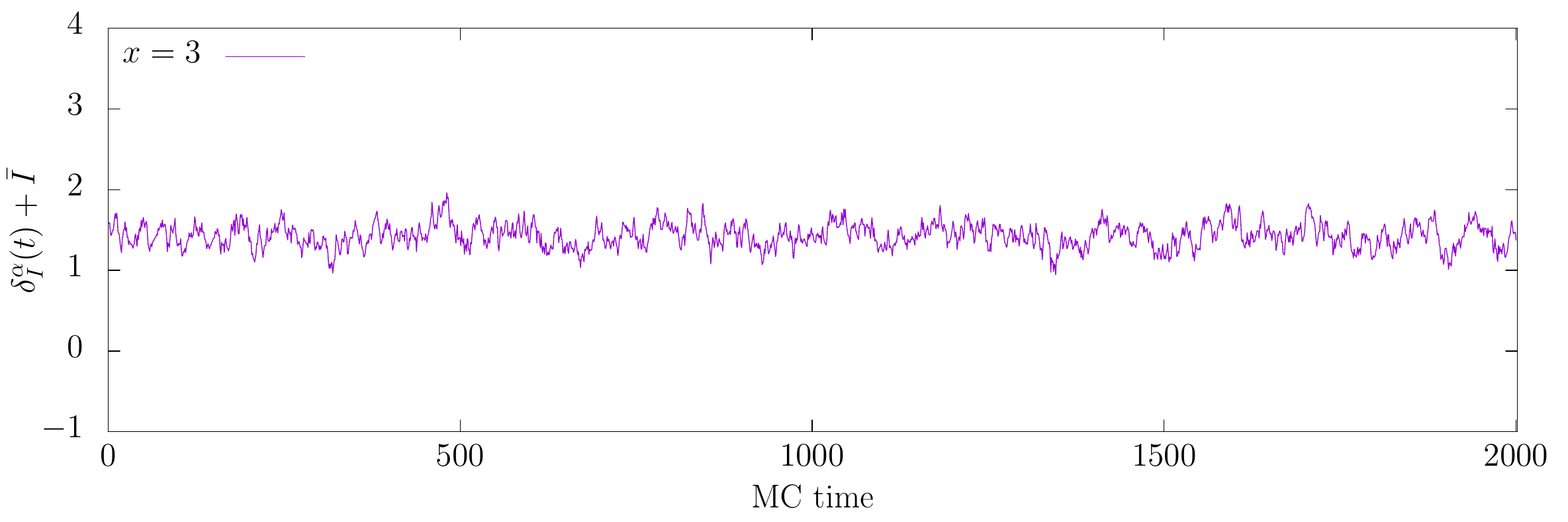}\\
  \includegraphics[width=0.95\textwidth,height=0.15\textheight]{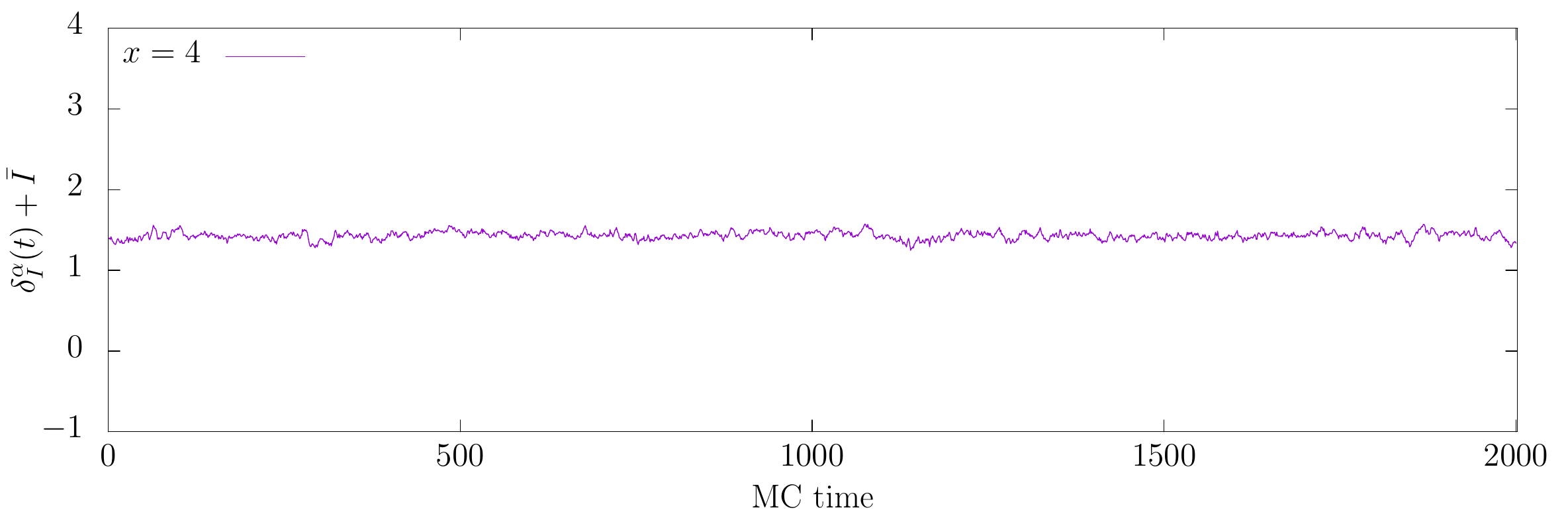}\\
  \includegraphics[width=0.95\textwidth,height=0.15\textheight]{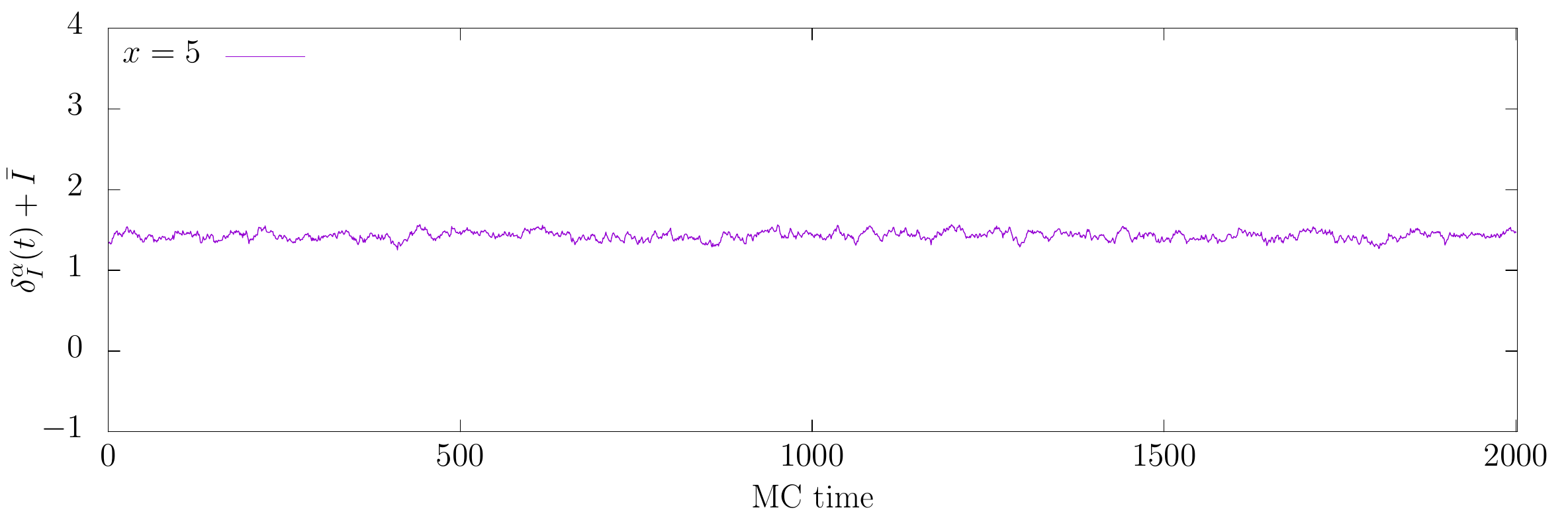}\\
  \includegraphics[width=0.95\textwidth,height=0.15\textheight]{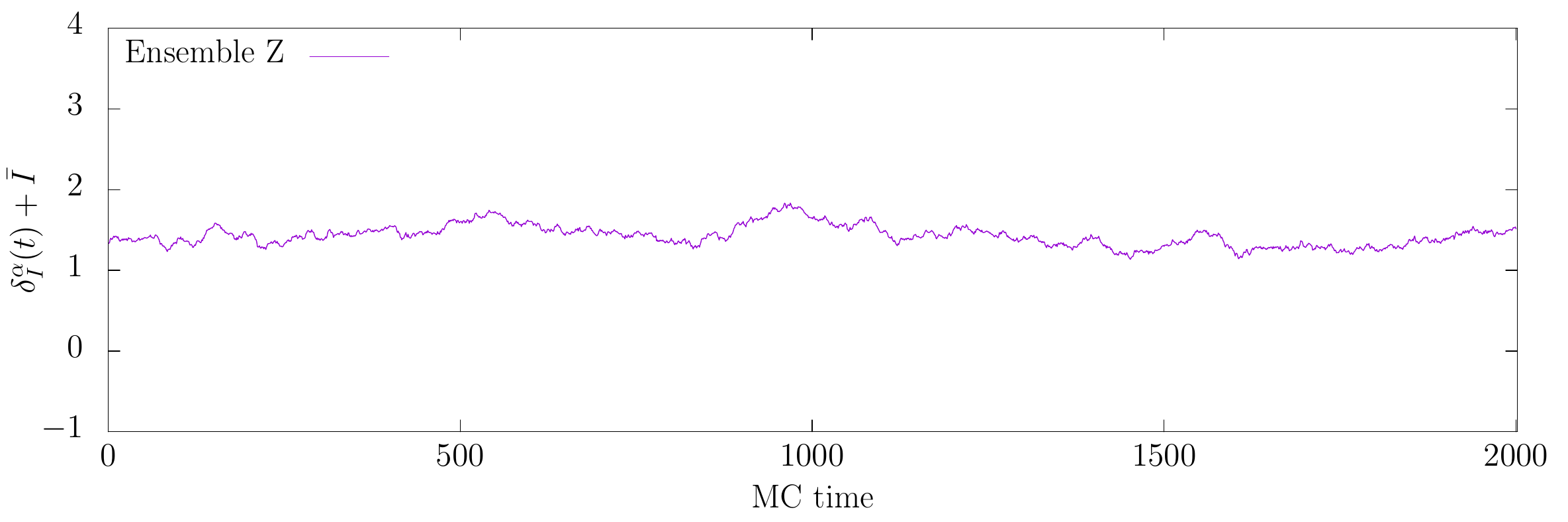}
  \caption{Fluctuations for the observable $I$ over the mean
    ($\delta_I^{\alpha}(t) + \bar I$) (see
    equation~(\ref{eq:error_propagation})) corresponding to 
    each ensemble that contributes to its error ($\alpha =
    x_i,Z$). Despite fluctations 
    being small in amplitude, ensemble $Z$ contributes a 61\% to the
    error in $I$ (the largest contribution).}
  \label{fig:hist}
\end{figure}

One of the advantages of the $\Gamma$-method is that the fluctuations
per ensemble are available even for complicated derived
observables. This gives access to the contribution of each ensemble to
the total error (see equation~(\ref{eq:error_contribution})), as well
as to the the fluctuations of the derived observable with respect to
any of the ensembles. Focusing our attention in observable $I$ we see 
(cf. table~\ref{tab:det}) that the largest 
contribution to the error comes from the observable $Z$, while
ensembles 3, 4 and 5 contribute very little. Figure~\ref{fig:hist}
shows the MC history of the fluctuations in $I$ for each Monte Carlo
ensemble. The main source of error, has
in fact fluctuations with a small amplitude, but the large
exponential autocorrelation time in this ensemble makes it the main
source of error in our determination of $I$. 

\begin{table}
  \small
  \centering
  \begin{tabular}{lllllll}
    \toprule
    &\multicolumn{6}{c}{Contribution to error} \\
    Quantity &  {Ensemble 1}& {Ensemble 2}   &
      {Ensemble 3}& {Ensemble 4}   &
                                     {Ensemble 5}& {Ensemble $Z$} \\
    \cmidrule(lr){1-1} \cmidrule(lr){2-7}
    $n$  & 35.08\,\% & 0.62\,\% & 1.29\,\% & 0.68\,\% & 22.07\,\%&40.27\,\% \\
    $m$  & 9.40\,\%  & 22.98\,\%  & 14.35\,\% & 0.04\,\% & 53.19\,\% & 0.04\,\% \\
    $I$  & 21.36\,\% & 11.89\,\% &  4.32\,\% & 0.72\,\% & 0.70\,\% & 61.01\,\% \\
    \bottomrule
  \end{tabular}
  \caption{Details of the analysis for the quantity that correspond to
  the interpolation of the fitted function at $x=3/2$. We observe that
  despite being an interpolation between the data of ensembles 1 and
  2, the largest contribution to the error comes from the parameter
  $Z$ (Note that $Z$ has the smaller relative uncertainty in
  table~\ref{tab:data}). }
  \label{tab:det}
\end{table}

\subsection{Exact error propagation}
\label{sec:exact-error-prop}

In order to support our claim that AD techniques perform linear
propagation of errors \emph{exactly}, it is interesting to
compare the result of the procedure
described in section~\ref{sec:fits} (based on computing the Hessian at
the minimum) with an implementation of the 
fitting procedure where all operations are performed in the hyper-dual
field. For this last case we use a simple (and inefficient) minimizer:
a simple gradient descent with a very small and constant damping
parameter. We point out that our inefficient
algorithm needs $\mathcal O(2000)$ iterations of the loop in
algorithm~\ref{alg:gradient} to converge to the solution. 

Since each operation in the algorithm is performed in the hyper-dual
field we are exactly evaluating the derivative of the
fitting routine with respect to the data points that enter in the
evaluation of the $\chi^2$. These derivatives are all that is needed to
perform linear error propagation. All derivatives are computed 
to machine precision.

Being more explicit, if we define a function that performs one iteration of the
gradient descent (see algorithm~\ref{alg:gradient})
\begin{equation}
  \mathbf{\mathcal I}(\mathbf{x}) = \mathbf{x} - \gamma\mathbf{\nabla}\chi^2\Big|_{\mathbf{x}}\,,
\end{equation}
we can define the function
\begin{equation}
  \mathcal M = \underbrace{\mathcal I \circ \mathcal I \circ \cdots
    \circ \mathcal I}_ {\text{2000 times!}}\,,
\end{equation}
that returns the minima of the $\chi^2$ for any reasonable input.
AD is computing the derivative of $\mathcal M$ with respect to the
data that enter the evaluation of the $\chi^2$
exactly\footnote{Incidentally it also computes the derivative with
  respect to the initial guess of the minima, and correctly gives zero.}. 

On the other hand the derivation of section~\ref{sec:fits} is also
exact to leading order. Therefore both 
procedures should give the same errors, even if in one case we are
computing the derivative just by making the derivative of every
operation of the gradient descent algorithm, and in the other case we
are exactly computing the Hessian of the $\chi^2$ function at the
minima and using it for linear error propagation. 
The results of this small test confirm our expectations: 
\begin{eqnarray}
\textbf{Gradient:}\quad n = 0.97959028626531608  &\pm&  0.193130534135\,80400\,.\\ 
\textbf{Hessian:}\quad  n = 0.97959028626531608  &\pm&  0.193130534135\,94974\,.\\ 
\nonumber  \\
\textbf{Gradient:}\quad m = 0.36501862665927676  &\pm&  4.4699145709\,887811\times 10^{-2}\,.\\
\textbf{Hessian:}\quad  m = 0.36501862665927676  &\pm&  4.4699145709\,956804\times 10^{-2}\,.
\end{eqnarray}
As the reader can see both errors agree with more than 12 decimal
places. The critical reader can still argue that in fact the errors
are not exactly the same. One might be tempted to say that the
small difference is due to ``higher orders terms'' in the expansion
performed in section~\ref{sec:fits},
but this would be wrong (AD is an \emph{exact} truncation up to some
order). The reason of the difference are ``lower order
terms''. In the derivation of section~\ref{sec:fits} we have assumed
that the $\chi^2$ is in the minimum. In fact, any minimization
algorithm returns the minimum only up to some precision. This
small deviation from the true minimum gives a residual gradient of the
$\chi^2$ function that contaminates (the 14$^{\underline{th}}$
significant digit!) of the errors in the fit parameters.  
\begin{algorithm}
  \caption{Gradient descent}
  \label{alg:gradient}
  \begin{algorithmic}[1]
    \State $\gamma \gets 0.0001$
    \State $X_i^{(1)} \gets 1.0$
    \State $\epsilon\gets 10^{-12}$
    
    \Repeat
    \State $X_i^{(0)} \gets X_i^{(1)}$
    \State $df_i \gets {\partial \chi^2}/{\partial x_i}$
    \State $X_i^{(1)} \gets X_i^{(0)} - \gamma * df_i$
    \Until ($|df|< \epsilon)\, \MyAnd\, (|X^{(1)}-X^{(0)}|<\epsilon)$
  \end{algorithmic}
\end{algorithm}


\section{Conclusions}
\label{sec:conclusions}

Lattice QCD is in a precision era. Input from lattice QCD is used to
challenge the Standard Model in several key areas like flavor physics,
CP violation or the anomalous moment of the muon to mention a few
examples. It is very likely that if new physics is discovered in the
next ten years, lattice QCD will be used as input. Error analysis of
lattice data is a key ingredient in the task of providing this
valuable input to the community. 

The state of the art lattice simulations that provide this important
input to the particle physics community require simulations at small
lattice spacings and large physical volumes. It is well-known that
these simulations in general (and specially if topology
freezing~\cite{DelDebbio:2004xh} plays a role) have large exponential
autocorrelation times. The field continues to push their simulations
to smaller and smaller lattice spacings to be able to simulate
relativistic \emph{charm} and \emph{bottom} quarks comfortably, making
the issue of large autocorrelation times more severe.

There are good theoretical and practical reasons to use the
$\Gamma$-method as tool for data analysis. Despite these arguments are
known for more than ten years (see~\cite{Wolff:2003sm}), most of the
treatment of autocorrelations in current state of the art computations
use binning techniques, known to underestimate the errors, specially
in the current situation of large autocorrelations. There are known
solutions to these issues: the
$\Gamma$-method~\cite{Madras1988,Wolff:2003sm, Schaefer:2010hu} allows
to study in detail the autocorrelations even of complicated derived
observables, and to include in the error estimates the effect of the
slow modes of the MC chain. The author does not know any other
analysis technique that allow to estimate statistical uncertainties
conservatively at the simulation parameters of current state of the
art lattice QCD simulations.

In this work we have considered the analysis of general observables
that depend on several MC simulations with the $\Gamma$-method. We
have shown that linear error propagation can be performed
\emph{exactly}, even in arbitrarily complicated observables defined
via iterative algorithms.  Thanks to automatic differentiation we only
need to extend the operations and evaluation of intrinsic fundamental
functions to the field of hyper-dual numbers. Moreover error
propagation in certain iterative procedures can be significantly
simplified. In particular we have examined in detail the interesting
case of fitting some Monte Carlo data and the case of error
propagation in the determination of the root of a non-linear
function. We have shown that error propagation in these cases only
require to use AD once the fit parameters or the root of the function
are known. Conveniently, AD techniques are not needed in the fitting
or root finding algorithms and one can rely on external libraries to
perform these tasks. By comparing these techniques with an
implementation of a fitting algorithm where all operations are
performed in the hyper-dual field we have explicitly checked that
error propagation is performed exactly. In summary, AD techniques in
conjunction with the $\Gamma$-method offer a flexible approach to
error propagation in general observables.

Analysis of Monte Carlo data along the lines proposed in this work is
\emph{robust} in the sense that if in any analysis the central values
of the parameters are computed correctly, the exact nature of the
truncation performed in AD guarantees that errors will be correctly
propagated.  Although the focus in this work has been on the
applications of AD to analysis of Lattice QCD data, the ideas
described here might find its way to other research areas.

Implementing the $\Gamma$-method for data analysis is usually
cumbersome: different MC chains have to be treated independently and
an efficient computation of the autocorrelation functions requires to
use the Fast Fourier Transform. We provide a portable, freely
available implementation of an analysis code that handles the analysis
of observables derived from measurements on any number of ensembles
and any number of replicas. Error propagation, even in iterative
algorithms, is exact thanks to AD~\cite{aderrors-mod}. We hope that
future analysis in the field can either use it directly, or as a
reference implementation for other robust and efficient analysis tools
of MC data.


\section*{Acknowledgments}
\addcontentsline{toc}{section}{Acknowledgments}

This work has a large debt to all the members of the ALPHA
collaboration and specially to the many discussions with Stefano
Lottini, Rainer Sommer and Francesco Virotta. The author thanks Rainer
Sommer for a critical reading of an earlier version of the manuscript,
and Patrick Fritzsch for his comments on the text.

\appendix

\section{A model to study analysis of autocorrelated data}
\label{sec:model}

In algorithms with detailed balance the autocorrelation function of
any observable can be written as a sum of decaying exponential
\begin{equation}
  \Gamma(t) = \sum_k \zeta_k e^{-|t|/\tau_k}\,,\qquad
  (\zeta_k > 0)\,,
  \label{eq:gamma_general}
\end{equation}
where the $\tau_k$ are related with the left eigenvectors of the
Markov operator. They are universal in the sense that they are a
property of the algorithm. Different observables decay with the same
values $\tau_k$. On the other hand the ``couplings'' $\zeta_k$ are
observable dependent. 

As proposed in~\cite{Wolff:2003sm} one can simulate noisy
autocorrelated data using Gaussian independent random numbers
\begin{equation}
\eta(t) \sim \mathcal N(0,1)\,,
\end{equation}
just defining
\begin{equation}
  \nu^{(k)}(t) = \sqrt{1-e^{-2/\tau_k}}\, \eta(t) + e^{-1/\tau_k} \nu(t-1)\,.
\end{equation}
The autocorrelation function for the variable $\nu^{(k)}(t)$ is
trivially computed to be $e^{-|t|/\tau_k}$. One can now construct a
linear combination of such variables 
\begin{equation}
  x(t) = x_{\rm mean} + \sum_k \lambda_k \nu^{(k)}(t)\,,
\end{equation}
that has an autocorrelation function of the type
equation~(\ref{eq:gamma_general}). A straightforward computation yields
\begin{eqnarray}
  \Gamma_x(t) &=& \langle x(0)x(t) \rangle = \sum_k \lambda_k^2
                  e^{-|t|/\tau_k}\,,\qquad \left( \Gamma_x(0) = \sum_k\lambda_k^2\right)\,. \\
  \rho_x(t) &=& \sum_k
                \frac{\lambda_k^2}{\Gamma_x(0)}e^{-|t|/\tau_k}\,.\\
  \tau_{{\rm int},x} &=& \frac{1}{2} + \sum_{t=1}^\infty \rho_x(t) =
  \frac{1}{2} + \sum_k
                         \frac{\lambda_k^2}{\Gamma_x(0)(e^{1/\tau_k}-1)}\,.
                         \label{eq:tauint_exact}
\end{eqnarray}
Finally the error for a sample of length $N$ is
\begin{equation}
  \label{eq:error}
  (\delta x)^2 = \frac{1}{N}\left( \Gamma_x(0) + 2 \sum_k
    \frac{\lambda_k^2}{e^{1/\tau_k} - 1} \right)\,.
\end{equation}
And if we decide to truncate the infinite sum in
eq.~(\ref{eq:tauint_exact}) with a window of size $W$, we get
\begin{equation}
  \label{eq:errorW}
  (\delta_W x)^2 = \frac{1}{N}\left( \Gamma_x(0) + 2 \sum_k
    \frac{\lambda_k^2(1-e^{-W/\tau_k})}{e^{1/\tau_k} - 1} \right)\,.
\end{equation}
Adding the tail with the slowest mode $\tau_{\rm exp} = \max\{\tau_k\}$
after summing the autocorrelation function up to $W$ gives as result
\begin{equation}
  \label{eq:errorWtexp}
  (\delta_{\rm exp} x)^2 = \frac{1}{N}\left( \Gamma_x(0) + 2 \sum_k
    \frac{\lambda_k^2(1-e^{-W/\tau_k})}{e^{1/\tau_k} - 1} + 2\tau_{\rm
    exp}e^{-(W+1)/\tau_{\rm exp}}\sum_k\lambda_k^2  \right)\,.
\end{equation}

\subsection{Binning}
\label{sec:ap_binning}

Binning can also be studied exactly in this model. Bins of length
$N_B$ are defined by block averaging
\begin{equation}
  b(\alpha) = \frac{1}{N_B}\sum_{i=1}^{N_B} x((\alpha-1)N_B + i)\,.
\end{equation}
The autocorrelation function of the binned data is given by
\begin{eqnarray}
  \Gamma_b(t) &=& \langle b(\alpha+t) b(\alpha) \rangle =
                  \frac{1}{N_B^2}\sum_{i,j=1}^{N_B} \Gamma_x(tN_B + i-j) \\
              &=& \frac{1-\delta_{t,0}}{N^2_B} \sum_k \lambda_k^2 K(N_B, \tau_k) +
                  \frac{\delta_{t,0}}{N^2_B} \sum_k \lambda_k^2 H(N_B,\tau_k)\,,
\end{eqnarray}
where the functions $K(N,\tau)$ and $H(N,\tau)$ are given by
\begin{eqnarray}
  K(N,\tau) &=& 
                \frac{1-\cosh(N/\tau)}{1-\cosh(1/\tau)}\,, \\
  H(N,\tau) &=& N\left[
    1+2e^{-1/\tau}\frac{1-e^{-(N-1)/\tau}}{1-e^{-1/\tau}}
                \right] \\
  \nonumber
  &-& 2e^{-1/\tau}\frac{1-Ne^{-(N-1)/\tau}+(N-1)e^{-N/\tau}}
  {(1-e^{-1/\tau})^2}\,. 
\end{eqnarray}
Now we can determine the normalized autocorrelation function and the
integrated autocorrelation time of the binned data
\begin{eqnarray}
  \Delta &=& \sum_k \lambda_k^2 H(N_B,\tau_k)\,, \\
  \rho_b(t) &=& \Delta^{-1}\sum_k\lambda_k^2
                K(N_B,\tau_k) e^{-tN_B/\tau_k}\,, \qquad (t>0)\,.\\
  \tau_{{\rm int},b} &=& \frac{1}{2} + \Delta^{-1} \sum_k \lambda_k^2
                         \frac{K(N_B,\tau_k)}{e^{N_B/\tau_k} - 1}\,.
\end{eqnarray}
Resampling techniques (bootstrap, jackknife) treat the bins as
independent variables. Therefore the error estimate for a sample of
length $N$ is
\begin{equation}
  (\delta_{\rm binning} x)^2 = \frac{\Delta}{NN_B^2}\,,
\end{equation}
while the true error eq.~(\ref{eq:error}) can conveniently be written as
\begin{equation}
  (\delta x)^2 = \frac{1}{NN_B^2}\left( \Delta + 2\sum_k \lambda_k^2
                         \frac{K(N_B,\tau_k)}{e^{N_B/\tau_k} - 1} \right)\,,
\end{equation}


\section{A free implementation of the $\Gamma$-method with AD error
  propagation} 
\label{sec:code}

Here we present a freely available \texttt{fortran 2008}
library for data analysis using the $\Gamma$-method with AD for linear error
propagation. The software is standard compliant and have no external
dependencies beyond a \texttt{fortran} compiler that supports the
modern standards\footnote{In particular, the code has been tested with
\texttt{gfortran} versions 6.X, 7.X, 8.X, \texttt{intel fortran}
compiler v17, v18 and the \texttt{cray} family of compilers.}. The
code can handle the analysis of several replica and observables from
different ensembles. For a detailed documentation and to obtain a copy
of the software check~\cite{aderrors-mod}.

\subsection{A complete example}

This is a complete commented example that shows most of the
features of the code. This particular snippet is part of
the distribution~\cite{aderrors-mod} (file \texttt{test/complete.f90}
(see~\ref{sec:code-listing})). It uses the 
\texttt{module simulator} (\texttt{test/simulator.f90}) that generates
autocorrelated data along the lines of appendix~\ref{sec:model}. Here
we give an overview on the package using a full example where we
compute the error in the derived quantity
\begin{equation}
  \label{eq:derived}
  z= \frac{\sin x}{\cos y + 1} \,.
\end{equation}
The quantities $x$ and $y$ are
generated using the procedure described in appendix~\ref{sec:model}
and are assumed to originate from simulations with
completely different parameters. The line numbers correspond to the
listing in section~\ref{sec:code-listing}.

\begin{figure}
  \centering
  \begin{subfigure}[b]{\textwidth}
    \includegraphics[width=0.9\textwidth]{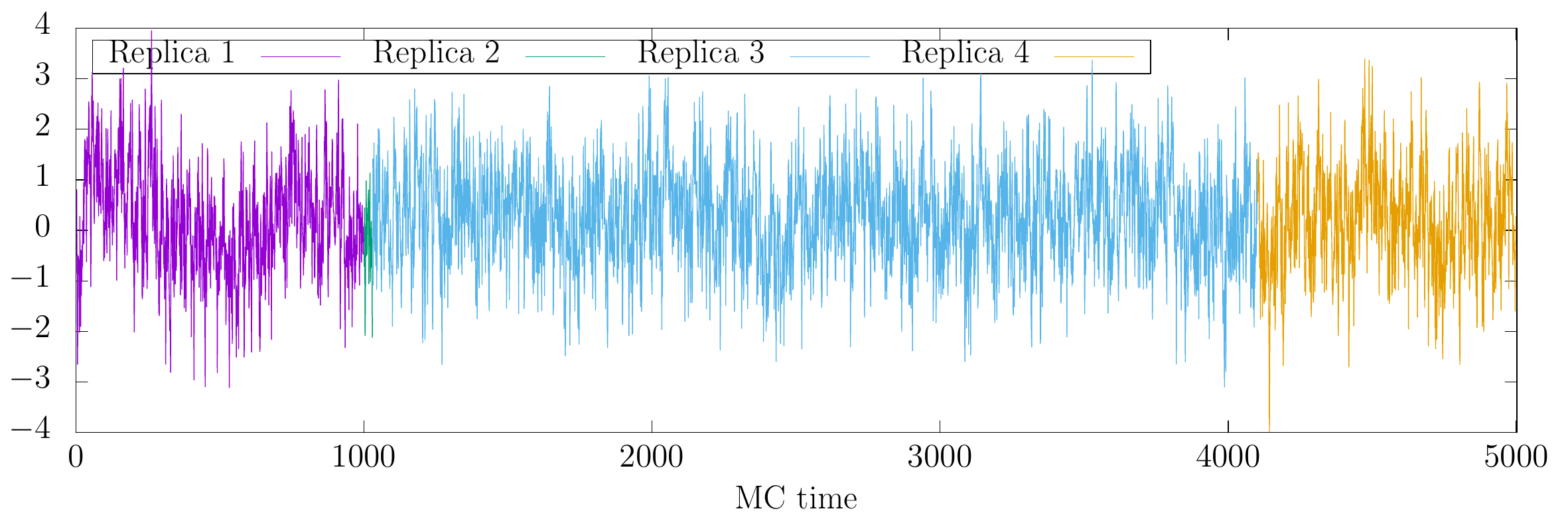}
    \caption{MC history for ID 1}
    \label{fig:h1}
  \end{subfigure}\\
  \begin{subfigure}[b]{0.45\textwidth}
    \includegraphics[width=\textwidth]{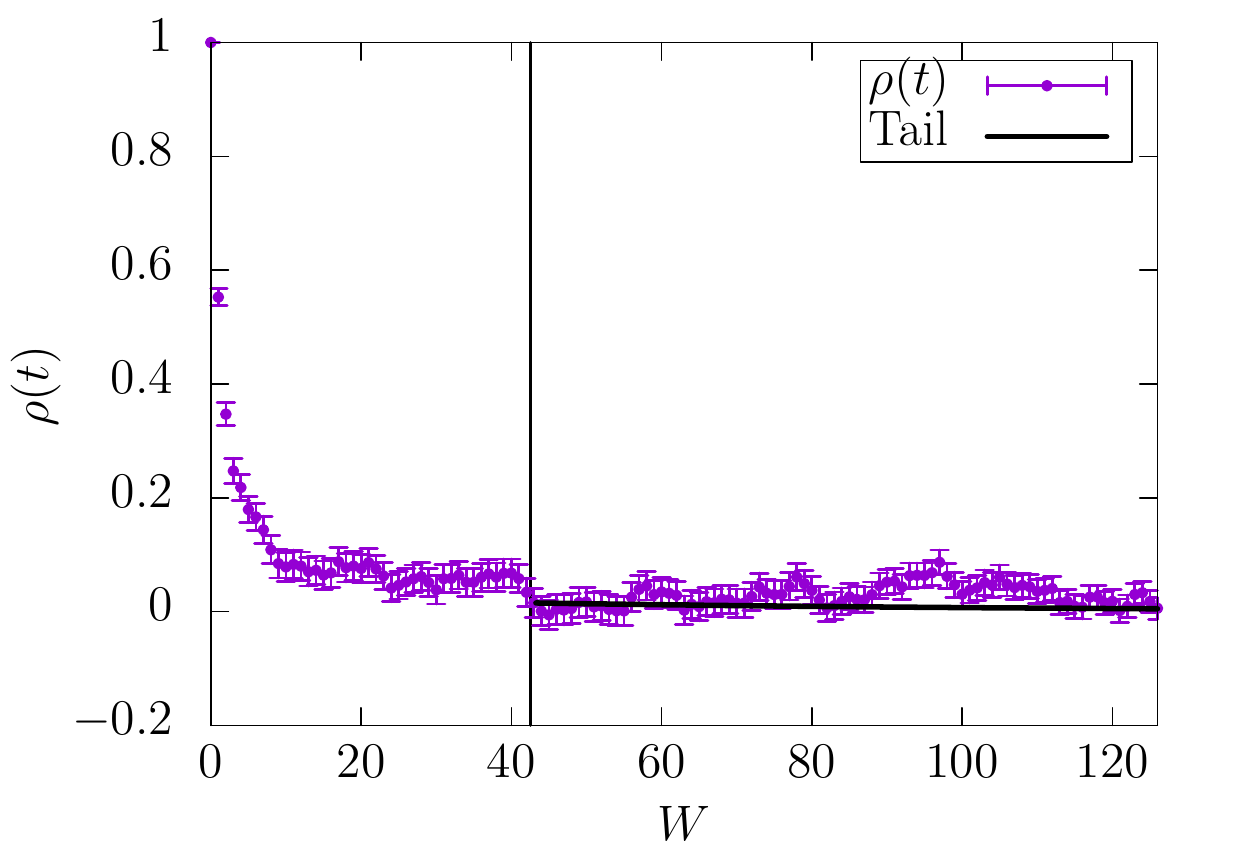}
    \caption{Normalized autocorrelation function for ID 1}
    \label{fig:a1}
  \end{subfigure}
  \begin{subfigure}[b]{0.45\textwidth}
    \includegraphics[width=\textwidth]{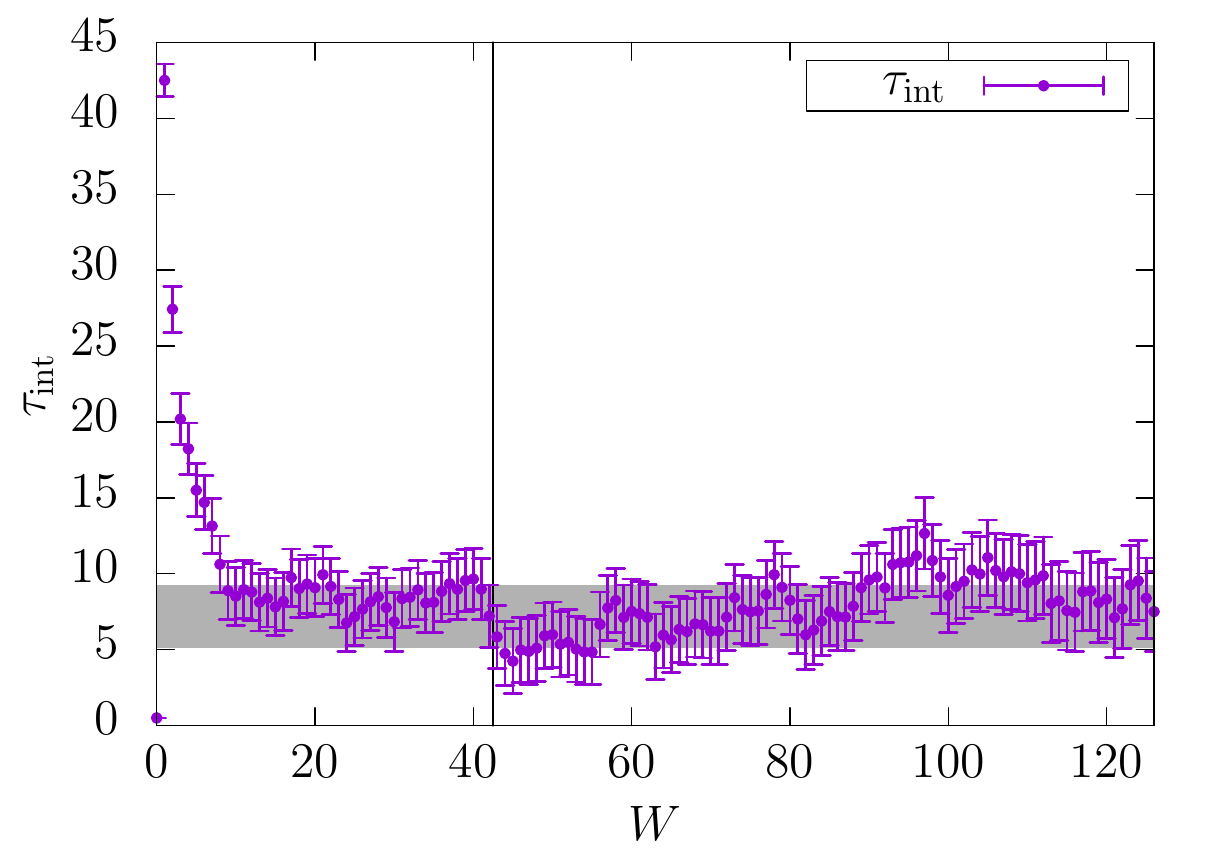}
    \caption{$\tau_{\rm int}$ for ID 1}
    \label{fig:t1}
  \end{subfigure}

  \begin{subfigure}[b]{\textwidth}
    \includegraphics[width=0.9\textwidth]{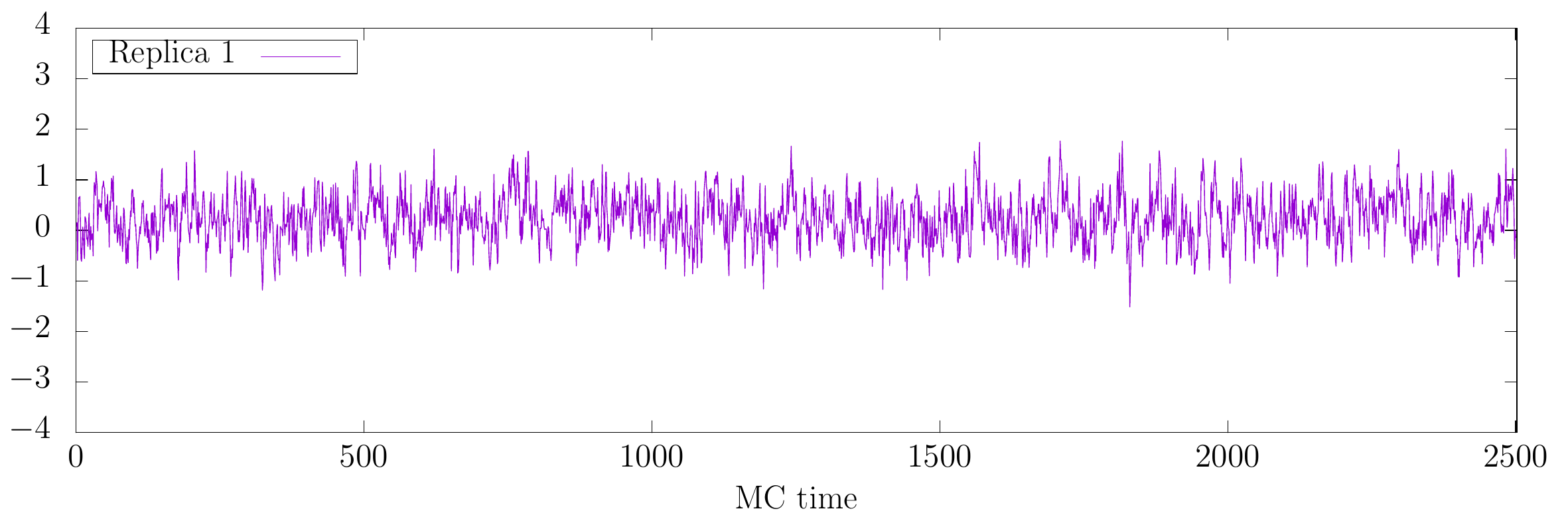}
    \caption{MC history for ID 2}
    \label{fig:h2}
  \end{subfigure}\\
  \begin{subfigure}[b]{0.45\textwidth}
    \includegraphics[width=\textwidth]{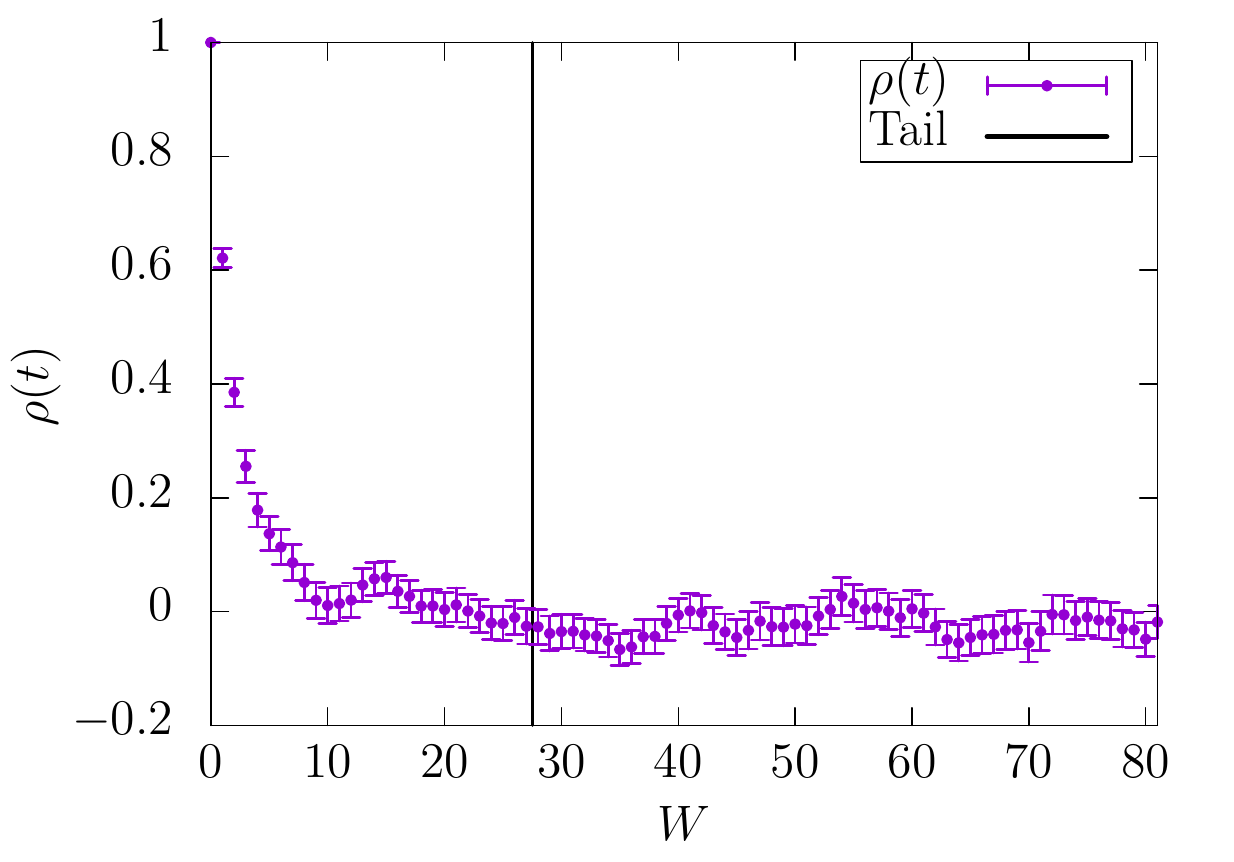}
    \caption{Normalized autocorrelation function for ID 2}
    \label{fig:a2}
  \end{subfigure}
  \begin{subfigure}[b]{0.45\textwidth}
    \includegraphics[width=\textwidth]{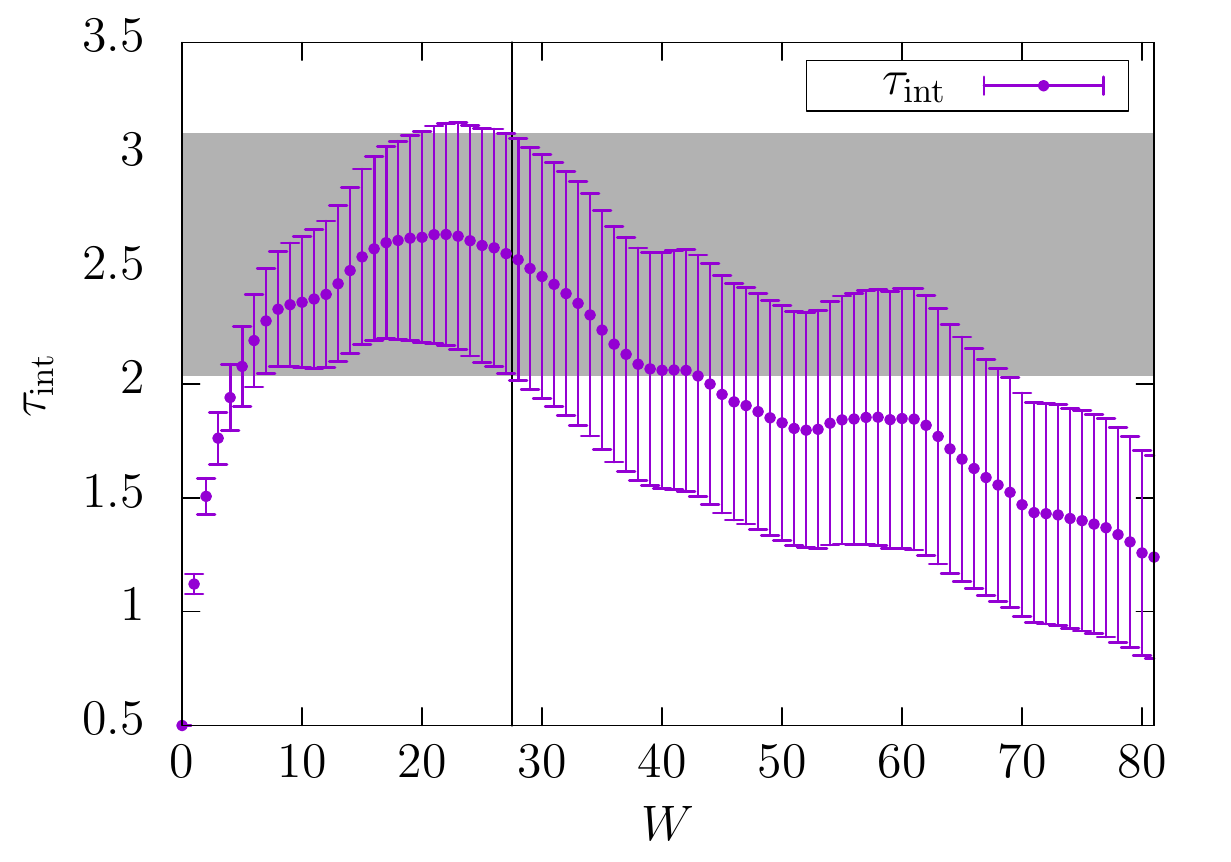}
    \caption{$\tau_{\rm int}$ for ID 2}
    \label{fig:t2}
  \end{subfigure}
  \caption{Histories, autocorrelation functions and $\tau_{\rm int}$
    for the derived observable $z$ (eq.~(\ref{eq:derived})).}
  \label{fig:hist_z}
\end{figure}

\begin{description}
\item[lines 4-7] modules provided with the
  distribution~\cite{aderrors-mod}. 
\item[lines 21-36] Use module \texttt{simulator} to produce measurements for
  two observables from different ensembles. On the first MC ensemble
  we have $\tau_k = [1.0, 3.0, 12.0, 75.0]$ and the measurements
  \texttt{data\_x(:)} correspond to couplings  
  $\lambda_k = [1.0, 0.70, 0.40, 0.40]$. For the second MC ensemble we
  have $\tau_k = [2.0, 4.0, 6.0, 8.0]$ and the measurements
  \texttt{data\_y(:)} have couplings $\lambda_k = [2.3, 0.40, 0.20,
  0.90]$. In the first case we have a sample of length 5000 in 4
  replicas of sizes $1000,30,3070,900$. For the second MC chain we
  have a single replica of size 2500. See 
  appendix~\ref{sec:model} for analytic expressions of the error of
  both observables. 

\item[lines 42-45] Load measurements of the first observable in the
  variable \texttt{x}, and set the details of the analysis:
  \begin{description}
  \item[line 43] Set the ensemble ID to 1.
  \item[line 44] Set replica vector to $[1000,30,3070,900]$.
  \item[line 45] Set the exponential autocorrelation time
    ($\tau_{\rm exp} = 75$).
  \end{description}
\item[lines 48-49] Load measurements of the second ensemble into
  \texttt{y} and set ensemble ID to 2. In this case we have only one
  replica (the default), and we choose not to add a tail to the
  autocorrelation function since the number of measurements (2500) is
  much larger than the exponential autocorrelation time $\tau_{\rm
    exp}=8$ of the second MC ensemble.

\item[line 52] Computes the derived observable \texttt{z}.

\item[lines 56-58] Details of the analysis for observable \texttt{z}:
  \begin{description}
  \item[line 56] Add the tail to the normalized autocorrelation
    function at the point where the signal $\rho(t)$ is equal
    to 1.0 times the error and the ensemble ID is 1.
  \item[line 58] Set the parameter $S_\tau=3$ for ensemble ID 2 to
    automatically choose the optimal window (see~\cite{Wolff:2003sm}).
  \end{description}

\item[line 60] Performs the error analysis in \texttt{z}.
  
\item[lines 62-68] Prints estimate of the derived observable with
  the error. Also prints $\tau_{\rm int}$ for each ensemble and what
  portion of the final error in \texttt{z} comes from each ensemble
  ID.

\item[lines 70-72] Prints in file \texttt{history\_z.log} the details
  of the analysis: fluctuations per ensemble, normalized
  autocorrelation function and $\tau_{\rm int}$ as a function of the
  Window size. This allows to produce the plots in Fig.~\ref{fig:hist_z}.
\end{description}

Running the code produces the output
\begin{verbatim}
 Observable z:     0.24426076465139626       +/-    5.8791563778643217E-002
   Contribution to error from ensemble ID  1   86.57%  (tau int: 7.2020 +/-   2.0520)
   Contribution to error from ensemble ID  2   13.43%  (tau int: 2.5724 +/-   0.5268)
\end{verbatim}

\newpage
\subsubsection{Example code listing}
\label{sec:code-listing}

\begin{lstlisting}
program complete

  use ISO_FORTRAN_ENV, Only : error_unit, output_unit
  use numtypes
  use constants
  use aderrors
  use simulator
  
  implicit none

  integer, parameter  :: nd = 5000, nrep=4
  type (uwreal)       :: x, y, z
  integer             :: iflog, ivrep(nrep)=(/1000,30,3070,900/), i, is, ie
  real (kind=DP)      :: data_x(nd), data_y(nd/2), err, ti, texp
  real (kind=DP)      :: tau(4), &
       lam_x(4)=(/1.0_DP, 0.70_DP, 0.40_DP, 0.40_DP/), &
       lam_y(4)=(/2.3_DP, 0.40_DP, 0.20_DP, 0.90_DP/)
  character (len=200) :: flog='history_z.log'


  ! Fill arrays data_x(:) with autocorrelated
  ! data from the module simulator. Use nrep replica
  tau = (/1.0_DP, 3.0_DP, 12.0_DP, 75.0_DP/)
  texp = maxval(tau)

  is = 1
  do i = 1, nrep
     ie = is + ivrep(i) - 1
     call gen_series(data_x(is:ie),  err, ti, tau, lam_x, 0.3_DP)
     is = ie + 1
  end do

  ! Fill data_y(:) with different values of tau also using
  ! module simulator
  forall (i=1:4) tau(i) = real(2*i,kind=DP)
  call gen_series(data_y, err, ti, tau, lam_y, 1.3_DP)


  ! Load data_x(:) measurements in variable x.
  ! Set replica vector, exponential autocorrelation time
  ! and ensemble ID.
  x = data_x
  call x%set_id(1)
  call x%set_replica(ivrep)
  call x%set_texp(texp)
  
  ! Load data_y(:) measurements in variable y
  y = data_y
  call y%set_id(2)

  ! Exact, transparent error propagation
  z = sin(x)/(cos(y) + 1.0_DP)

  ! Attach tail in ensemble with ID 1 when signal in the
  ! normalized auto-correlation function equals its error
  call z%set_dsig(1.0_DP,1) 
  ! Set Stau=3 for automatic window in ensemble with ID 2
  call z%set_stau(3.0_DP,2)
  ! Perform error analysis (tails, optimal window,...)
  call z%uwerr() 

  ! Print results and output details to flog  
  write(*,'(1A,1F8.5,1A,1F8.5)')'** Observable z:   ', z%value(),  " +/- ", z%error()
  do i = 1, z%neid()
     write(*,'(3X,1A,1I3,3X,1F5.2,"%")',advance="no")'Contribution to error from ensemble ID', &
          z%eid(i), 100.0_DP*z%error_src(i)
     write(*,'(2X,1A,1F0.4,1A,1F8.4,1A)')'(tau int: ', z%taui(i), " +/- ", z%dtaui(i), ")"
  end do

  open(newunit=iflog, file=trim(flog))
  call z%print_hist(iflog)
  close(iflog)
  
  stop
end program complete


\end{lstlisting}


\clearpage
\addcontentsline{toc}{section}{References}
\bibliography{/home/alberto/docs/bib/math,/home/alberto/docs/bib/campos,/home/alberto/docs/bib/fisica,/home/alberto/docs/bib/computing}

\end{document}